\documentclass{article}
\usepackage{graphicx,epsfig}
\usepackage{graphicx,subfigure}
\usepackage{lipsum}
\usepackage{arxiv}
\usepackage{amsmath}
\usepackage[utf8]{inputenc} 
\usepackage[T1]{fontenc}    
\usepackage{hyperref}       
\usepackage{url}            
\usepackage{booktabs}       
\usepackage{amsfonts,bm}       
\usepackage{microtype}      
\usepackage[mathscr]{euscript}
\usepackage[numbers]{natbib}
\title{Impact of sloping porous seabed on the efficiency of an OWC against oblique waves.}

\author{
  Mohamin B M Khan \\
  Department of Mathematics\\
  SRM Institute of Science and Technology\\
  Chennai, India \\
  \texttt{muhaymin77@gmail.com} \\
   \And
 Harekrushna Behera \thanks{Corresponding Author} \\
  Department of Mathematics\\
  SRM Institute of Science and Technology\\
  Chennai, India \\
  \texttt{hkb.math@gmail.com} \\
}

\begin{document}
\maketitle

\begin{abstract}
The study examines the influence of a sloping porous bed on the efficiency of an oscillating water column (OWC) device facing oblique water waves. A vertical, surface piercing, thin plate near a rigid wall approximates the OWC. The system is simulated using a multi-domain boundary element method assuming the linear potential theory. The impact of varying sloping bed structural parameters and influence of the incident wave angle on the performance of the OWC is evaluated and discussed. The significance of this model to act as a breakwater protecting near-shore marine facilities is also highlighted. The OWC efficiency is found to be highly sensitive to the slope of the porous seabed, and seabed porosity is found to stabilize the resonant frequency against changes in water levels. OWCs are found to be more efficient over porous seabeds with higher frictional coefficients. The results indicate optimum values for the parameters governing such a wave energy conversion system that can be used for the design and implementation of the model.
\end{abstract}


\section{Introduction}
Over recent decades, global renewable energy enterprise has recognized the potential of ocean waves for energy generation. The estimated amount of worldwide annual wave energy by The World Energy Council is 17.5 PWh (PWh: Peta-Watt hours = $10^{12}$ kWh), which is comparable with the estimated annual worldwide energy consumption of 16 PWh (\citet{boyle1996renewable}). Wave energy converter technology is paving the way for increasingly efficient means of ocean energy extraction. In particular, oscillating water columns in various configurations are the most advantageous and worthwhile choices for extensive research in this area.  OWC consists of open end box made of concrete or steel partly immersed in the ocean. The maximum power absorption by the OWC is attained when the resonance frequency of the water column inside the chamber is equal to the frequency of the incoming waves. The advantages of the OWC devices with attached air turbine stem from their reliability, efficiency, and low cost of maintenance. On a practical level, they have very few moving parts, and none are in the water. The concept is adaptable and can be used on a range of collector forms situated on the coastline, in the near-shore region, or floating offshore.   

In the recent decades, most of the R$\&$D activity in wave energy has been taking place in Europe, largely due to the financial support and coordination provided by the European Commission, and to the positive attitude adopted by some European national governments. Simultaneously, growing interest in wave energy is taking place in northern America (USA and Canada), involving the national and regional administrations, research institutions and companies, and giving rise to frequent meetings and conferences on ocean energy (\citet{antonio2010wave}). In Japan, near the port of Sakata, the  wave energy converter was integrated with a breakwater having rating 60 kW and became operational in 1989 (\citet{heath2012review}). Subsequently, a single turbine 75 kW OWC unit was built by Queen’s University Belfast on Islay, Scotland following which they designed the LIMPET plant near to the original plant rated at 500 kW and has been in operation since late 2000. At the same time as the LIMPET development, a 400 kW OWC was built by Electricidade dos Açores on the island of Pico. The most recent OWC deployment, developed by Oceanlinx, involved its MK3 floating device which was a one-third scale demonstration version of the 2.5 MW full-scale device. It was installed offshore from the eastern breakwater of Port Kembla Harbour from February to May 2010 (\citet{heath2012review}).

Early theories of wave energy absorbers were concentrated on rigid body models for wave power extraction from water waves as described by \citet{evans1976theory}, \citet{evans1981power} and \citet{mei1976power}. Subsequently, \citet{evans1978oscillating}, extended these theories for the simple OWC by assuming a simple rigid floating body inside the chamber on free surface and the spring-dashpot system was used as the power take-off model. \citet{evans1982wave} further generalized the work by considering an arbitrary number of fixed chambers enclosing an
internal free surface, which act as a system of oscillating water columns. \citet{falnes1985surface} further extended that work by designing a wave energy absorption system consisting of any number of oscillating bodies and internal pressure distributions.  Later, \citet{evans1995hydrodynamic} by utilizing a Galerkin method investigated the wave energy device approximated by a vertical surface-piercing plate near a rigid wall.  By employing the method of eigenfunction expansion, \citet{deng2013wave} investigated the water wave interacting with a circular oscillating water column. The mentioned papers are the fundamental results for developing general theories for oscillating pressure systems.

Increasing the performance of OWC devices is one of the important aspects for commercialization of the device. In this regard, the maximum power absorption by the OWC is attained when the resonance frequency of the water column inside the chamber is equal to the frequency of the incoming waves. Based on this fact, the results are studied to indicate optimum values for the proposed system under prevalent environmental and geographic conditions, in order to maximize energy conversion efficiency. Additionally, various proposed structural and hydrodynamic conditions are analyzed to increase the performance of OWC devices. The roles of introducing stepped bottom, multi-chamber OWCs, undulated slope, a harbour in front of the OWC or a reflecting wall etc. are also studied with the aim of increasing performance of the systems (\citet{rezanejad2013stepped} , \citet{rezanejad2015analytical}). Recently, \citet{xu2019review} provided a review on floating wave energy converter and a procedure for mooring design.

In this paper, the interaction of OWC  with oblique ocean waves is simulated for a vertical floating plate near a rigid wall over a sloping porous bed. Moreover, this model is also implemented as a  breakwater system protecting the near-shore marine facilities. The mathematical problem is formulated in the two dimensional Cartesian coordinate system under the linear water wave theory. Different mathematical approaches are adopted to solve the associated boundary value problem. Modelling of the OWC systems and hydrodynamic analysis of wave and structural parameters is done using ca combination of matched eigenfunction expansion and multi-domain Boundary Element Method (BEM) using constant or linear elements. 

\section{Mathematical formulation}
\begin{figure}[ht!]
	\centering
	\subfigure[3D rendition of the fluid domain.]{\label{fig:1a}\includegraphics[trim={3in 0in 0in 0in},width=0.8\textwidth]{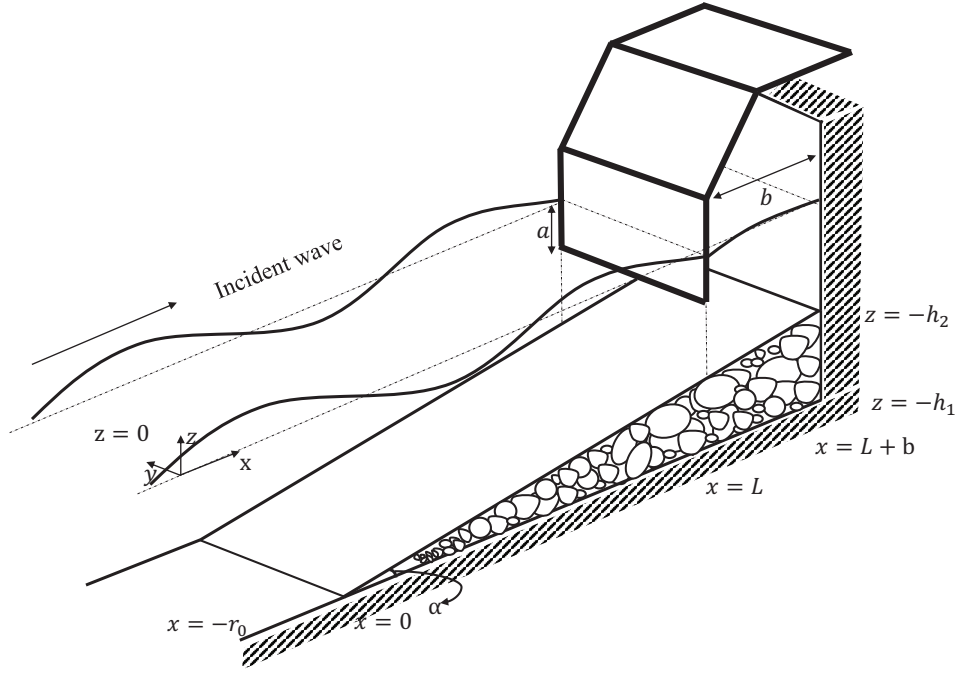}}
	\subfigure[BEM schematic for division of fluid domain.]{\label{fig:1b}\includegraphics[width=0.8\textwidth]{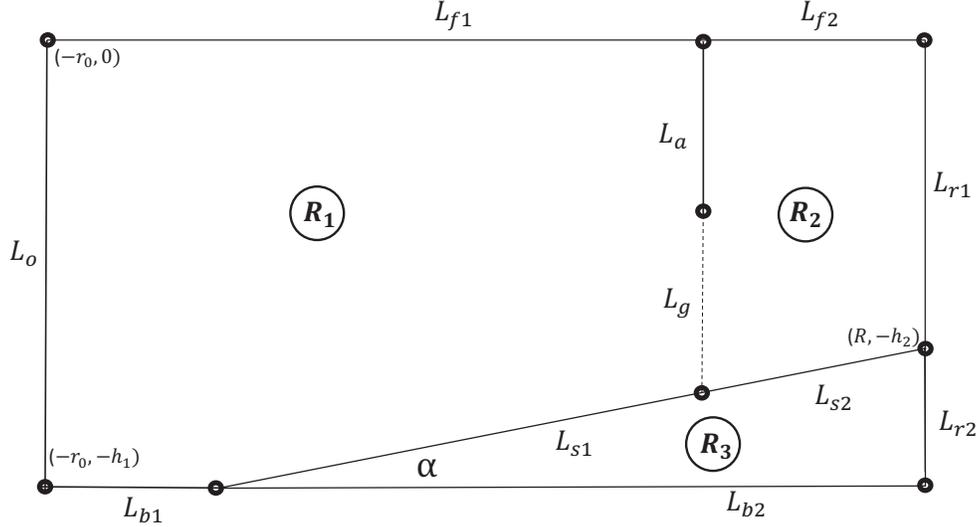}}
	\caption{Schematic diagram of surface-piercing wave energy converter with sloping porous sea bed.}
	\label{f1}
\end{figure}

The physical model is formulated in a  three-dimensional Cartesian coordinate system $(x,y,z)$ with $x-y$ plane parallel to the free surface and positive $z$ direction vertically upward. The OWC device is modeled by a thin surface piercing vertical plate near a rigid wall at $x=L$, while the bottom is covered with a sloping porous medium. The defining structural parameters of the system are submerged length of the vertical plate, distance from the rigid wall, depth of open water, depth of the rigid wall and the sloping angle of the seabed, denoted by $a,b,h_1,h_2$ and $\alpha$ respectively (as in Fig.\ref{fig:1a}).

The fluid properties are characterized by the scalar velocity potential $\Phi(x,y,z,t)$ satisfying $\nabla^2\Phi=0$, under assumptions of incompressibility, inviscid, and irrotational flow. The assumptions of  infinite length in $y-$ direction and time harmonic motion of waves with angular velocity $\omega$ extract $y$ and $t$ dependence such that $k_y = k_0 \sin\theta$ where $\theta$ is the incident wave angle, $k_0$ is the progressive  wave number, and $k_0$ satisfies the dispersion relation $\omega^2 = gh\tanh(kh)$. The existing velocity potential $\Phi(x,z,t)=Re\{\varphi(x,z)e^{-\mathrm{i}(k_yy-\omega t)}\}$, thus, satisfies the Helmholtz equation  
\begin{align}
	(\partial_{xx} +\partial_{zz} - k_y^2)\varphi=0. \label{helm}
\end{align}
The multi-domain boundary element method solution requires an auxiliary boundary at $x = -r_0$ and the resulting flow domain is divided into three regions, $R_1,R_2$ and $R_3$,  using a virtual boundary $L_g$ at $x=L$ as shown in Fig \ref{fig:1b}. The corresponding boundary conditions are given as: 
\begin{align}
	&\text{Rigid boundary condition: }&\partial&_x \varphi=0, \quad \mbox{at} \quad L_{b1},L_{b2}, L_{r1}, L_{r2},\label{bc1}\\
	&\text{No flow condition: }&\partial&_x \varphi=0, \quad \mbox{at} \quad L_{a},\\
	&\text{Velocity and pressure continuity I: } &\varphi&|_{x=b^-}=\varphi|_{x=b^+}, \quad  \partial_x\varphi|_{x=b^-}=\partial_x\varphi|_{x=b^+} \quad \mbox{at} \quad L_g,\label{bc2}\\
	&\text{Velocity and pressure continuity II: } &\varphi&_{1,2}= 
	(s-if)\varphi_3,\quad\displaystyle \partial_x{\varphi_{1,2}} = 
	\epsilon\partial_x{\varphi_3} \quad \mbox{at} \quad L_{s1,s2},\\
	&\text{Linearized free surface (external): }&\partial&_x\varphi+K\varphi=0 \quad \mbox{at} \quad L_{f1},\label{bc3}\\
	&\text{Linearized free surface (internal): }&\partial&_x\varphi+K\varphi= -\frac{\mathrm{i}\omega p}{\rho g} \quad \mbox{at} \quad L_{f2},\label{bclast}
\end{align}
where $K=\omega^2/g$, and $s,\epsilon,f$ are the inertial coefficient, porosity, and  frictional coefficient for the seabed, respectively. The distribution of pressure over the internal free-surface takes the form $P(t)=Re\{pe^{-\mathrm{i}\omega t}\}$. 
The total velocity potential is given by $\displaystyle \varphi=\varphi^S-\frac{\mathrm{i}\omega p}{\rho g}\varphi^R$ where $\varphi^S$ and $\varphi^R$ denotes the scattered and radiated potential. 
Consequently, the far-field condition is given by 
\begin{align}
	&\varphi^R(x,z)=A_R e^{\mathrm{i}\kappa(x-b)}\psi_{0}(k_0,z), \quad x\rightarrow -\infty , \\
	&\varphi^S(x,z)= (A_I e^{-\mathrm{i}\kappa(x-b)}+Re^{\mathrm{i}\kappa(x-b)})\psi_{0}(k_0,z), \quad x\rightarrow -\infty \label{13},
\end{align}
where $A_R, A_I$ are the amplitudes of radiated and incident waves,  $R$ is the unknown constant parameterizing wave reflection.  $\psi_0(k_0,z)$ is the eigenfunction in open water outside the problem domain and $\kappa=\sqrt{k_0^2-k_y^2}$.
The volume flux across the free surface $Q(t)=\text{Re}\{qe^{-\mathrm{i} \omega t}\}$ is as follows
\begin{align}
	q=\int_{L_{f2}}\partial_z \varphi~dx=q^S-\frac{\mathrm{i} \omega p}{\rho g}q^R, \text{ where } \frac{\mathrm{i}\omega p}{\rho g}q^R=-(\tilde{\mathcal{B}}-\mathrm{i}\tilde{\mathcal{A}}),\label{proc1}
\end{align}
with $q^S$ and $q^R$ being the volume fluxes across the internal free-surface $L_{f2}$. $\tilde{\mathcal{A}}$ and $\tilde{\mathcal{B}}$ are analogous to the added mass and radiation damping coefficient in a rigid body system, called as radiation susceptance and conductance parameters, respectively, given by 
\begin{align}
	\tilde{\mathcal{A}}= \frac{\omega}{\rho g}\text{Re}\{q^R\},\quad \tilde{\mathcal{B}}= \frac{\omega}{\rho g}\text{Im}\{q^R\}.
\end{align}
The volume flux through the turbine linearly proportional to the pressure drop across the internal free surface 
\begin{align}
	q=\upsilon p,
\end{align}
where $\upsilon$ is real. The mean rate of work done by the pressure force over one wave period is given by 
\begin{align}
	\mathcal{W}=\frac{|q_s|^2}{8\tilde{\mathcal{B}}} \Bigg[\frac{4\upsilon \tilde{\mathcal{B}}}{(\upsilon+\tilde{\mathcal{B}})^2+\tilde{\mathcal{A}}^2}\Bigg].
\end{align}
For known values of $\tilde{\mathcal{A}}$ and $\tilde{\mathcal{B}}$, the optimum value of $\upsilon$ is attained by  
\begin{align}
	\upsilon_{opt}^2=\sqrt{(\tilde{\mathcal{A}}+\tilde{\mathcal{B}})},
\end{align}
and thus the maximum work done by the pressure force over one wave period is 
\begin{align}
	\mathcal{W}_{max}=\frac{|q_s|^2}{8\tilde{\mathcal{B}}} \Bigg[\frac{2\tilde{\mathcal{B}}}{\upsilon_{opt}+\tilde{\mathcal{B}}}\Bigg].
\end{align}
The available power over one wave period of a plane progressive wave of unit amplitude is given by 
\begin{align}
	P_\mathcal{W}=Ec_g,
\end{align}
where $E$, $c_g$ are the total energy and group velocity given by 
\begin{align}
	E=\frac{\rho g}{2}, \quad c_g=\frac{\omega}{2k}\frac{1}{2}\Bigg(1+\frac{ 2kh_1}{\sinh 2kh_1}\Bigg).
\end{align}
The OWC efficiency which read as 
\begin{align}
	\eta_{max}=\frac{\mathcal{W}_{max}}{P_\mathcal{W}}.
\end{align}
%
The radiation susceptance and conductance are represented by non dimensional parameters $\mu$ and $\nu$ as (as in \citet{evans1995hydrodynamic})
\begin{align}
	\mu= \frac{\rho g}{\omega b}\tilde{\mathcal{A}},\quad \nu= \frac{\rho g}{\omega b}\tilde{\mathcal{B}}.\label{procl}
\end{align}
\section{Method of solution} \label{m2}\vspace*{-2mm}
The above boundary value problem is solved using the multi-domain boundary element method for the domains $R_1, R_2$ and $R_3$. Green's integral theorem is applied to Eq. \eqref{helm} using Green's function $G$ (\citet{khan2020analysis}), the resulting integral equation for a closed boundary $\Gamma$ is given as
\begin{equation}\label{intgeqn}
	-\left(\begin{array}{c} \varphi(x,z)
		\\
		\frac{1}{2}\varphi(x,z)
	\end{array}\right)
	=\int_{\Gamma}\Big(\varphi\partial_{\bm{n}}G-G\partial_{\bm{n}}\varphi\Big)\,
	d\Gamma,\,\left(\begin{array}{c}
		\mbox{if}\;(x,z)\,\in \mbox{int}(\Gamma)
		\\
		\mbox{if}\;(x,z)\;\in\; \Gamma 
	\end{array}\right),
\end{equation} 
After implementing the boundary conditions \eqref{bc1}--\eqref{bclast} into Eq. \eqref{intgeqn} for the elements on discretized boundaries of each region, along with the assumption of constant potential over each boundary element, a system of equations is obtained as
\begin{align}
	\begin{split}
		&C\varphi_1 +\int_{L_{f2}}\varphi_1\bigg(G_{\bm{n}}-KG\bigg)dL+\int_{L_O}\bigg(\varphi_1G_{\bm{n}}-G\varphi_{1,\bm{n}}\bigg)dL+\int_{L_{b1}} \varphi_1G_{\bm{n}}dL
		\\&+\int_{L_{s1}}\bigg(\varphi_1G_{\bm{n}}-G\varphi_{1,\bm{n}}\bigg)dL +\int_{L_{g}}\bigg(\varphi_1G_{\bm{n}}-G\varphi_{1,\bm{n}}\bigg)dL
		+\int_{L_{a}}\varphi_1G_{\bm{n}}dL~\qquad=\quad0,
	\end{split}\label{intra}
\end{align}
\begin{align}
	\begin{split}
		&C\varphi_2 +\int_{L_{f1}}\varphi_2\bigg(G_{\bm{n}}-KG\bigg)dL
		+\int_{L_{a}}\varphi_2G_{\bm{n}}dL
		+\int_{L_{g}}\bigg(\varphi_2G_{\bm{n}}-G\varphi_{2,\bm{n}}\bigg)dL
		\\&+\int_{L_{s2}}\bigg(\varphi_2G_{\bm{n}}-G\varphi_{2,\bm{n}}\bigg)dL
		+\int_{L_{r1}} \varphi_1G_{\bm{n}}dL~\qquad=\quad0,
	\end{split}\label{intra2}
\end{align}
\begin{align}
	\begin{split}
		&C\varphi_3 +\int_{L_{s1}}\bigg(\varphi_3G_{\bm{n}}-G\varphi_{3,\bm{n}}\bigg)dL
		+\int_{L_{b2}+L_{r2}}\varphi_3G_{\bm{n}}dL
		\\&+\int_{L_{s2}}\bigg(\varphi_3G_{\bm{n}}-G\varphi_{3,\bm{n}}\bigg)dL~\qquad=\quad0.
	\end{split}\label{intra3}
\end{align}

The above system is discretized over each boundary resulting in a matrix equation $\displaystyle\boldsymbol{H}\boldsymbol{U}=\boldsymbol{G}\boldsymbol{Q}+\boldsymbol{B}$, where
\begin{align}\displaystyle
	&\boldsymbol{H}=\nonumber\\&\left[ {\begin{array}{ccccccccccccccc}
			I(\varphi(z)) & H^1_{b1}& H^1_{s1}& H^1_{g}& H^1_a&(H^0 - KG^0)_{f}  & 0&0 &0&0&0&0\\
			0&     0   & 0& H^2_{g}&0&0&H^2_a  & H^2_{s2}& H^2_{r1} &(H^2 - KG^2)_{f}&0&0 \\
			0& 0& \frac{1}{(s-if)}H^3_{s1} &  0& 0& 0&0& \frac{1}{(s-if)}H^3_{s2}&0&0&H^3_{b2} & H^3_{r2}
	\end{array} } \right],\\
	&\boldsymbol{U}= \left[ {\begin{array}{ccccccccccccc}
			U^1_{O} & U^1_{b1} & U^1_{s1}& U^1_{g}& U^1_{a} & U^1_{f2}&U^2_{a}& U^2_{s2}& U^1_{l_2}&U^2_{r1}&U^2_{f1}&U^3_{b2}&U^3_{r2} 
	\end{array} } \right]^T,\\
	&\boldsymbol{G}=\left[ {\begin{array}{cccc}
			&G^1_{s1}&  G^1_g &0 \\
			&0& -G^2_g&G^2_{s2} \\
			&\frac{-1}{\epsilon}G^2_{s1} &  0&\frac{-1}{\epsilon}G^3_{s2}
	\end{array} } \right],\\
	&\boldsymbol{Q} =  \left[ {\begin{array}{ccc}
			Q^1_{s1}& Q^1_{g}& Q^2_{s2} 
	\end{array} } \right]^T,\\
	&\boldsymbol{B}=\left[ {\begin{array}{ccccccccccccccc}
			-2\mathrm{i}\kappa A_I\psi_0& 0 & 0& 0& 0&0& 0&0& 0&0&0&0&0&0&-\frac{\mathrm{i}\omega p}{\rho g}   
	\end{array} } \right]^T.
\end{align}
\begin{eqnarray}
	\text{Here, }H^i_j=C\delta_{ij}+\int_{L_{j}}\partial_{\bm{n}}G\;d
	L,\quad G^i_j=\int_{L_{j}}G\;dL, \quad U^i_j = \varphi^i_j \quad\text{and}\quad Q^i_j = \partial_{\bm{n}}\varphi^i_j,\quad \text{for}\quad R_i \text{ and }L_j.
\end{eqnarray}
The terms $I(\varphi(z))$ and 
$-2\mathrm{i}\kappa A_I\psi_0$ are evaluated using orthogonality of eigenfunctions exactly as in \citet{khan2020analysis}, such that the outside open water region has $\partial_x\varphi = I(\varphi(z)) -2\mathrm{i}\kappa A_I\psi_0$ at $L_O$. In $\boldsymbol{B}$, the first and last terms correspond to scattered and radiated potential respectively. The solution to this system of equations yields the values of potential and fluxes at the corresponding boundaries. The resulting values are processed using formulas in Eqs. \eqref{proc1}-\eqref{procl} and the following analysis is conducted. 
\begin{figure}[h!]
	\begin{center}
		\subfigure[$h_1=h_2=4$ m$,\frac{b}{h_1}=1,\theta =0^{\circ}$]{\label{fig:va}\includegraphics*[width=0.49\textwidth]{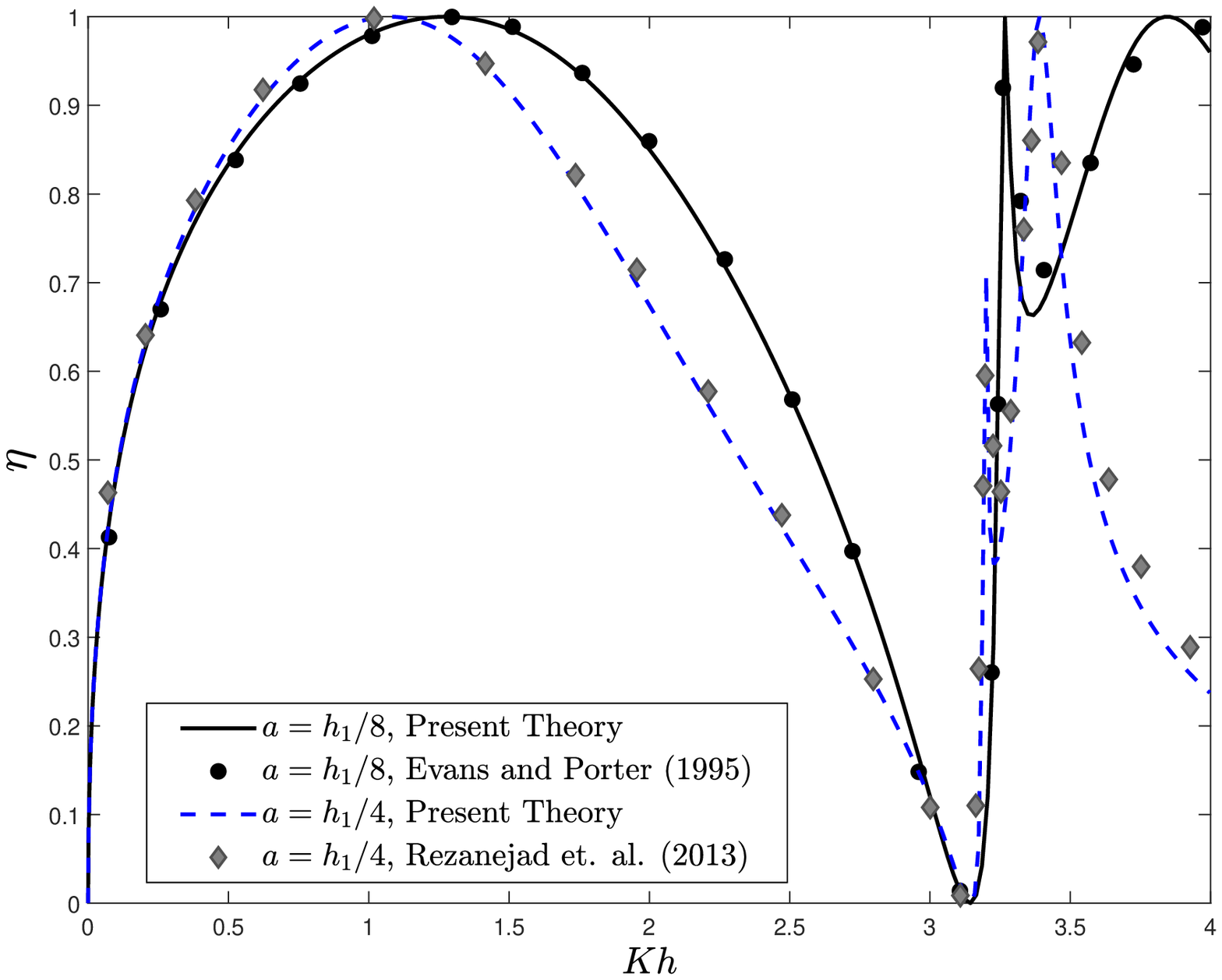}}
		\subfigure[$h_1=4$ m$, \frac{h_2}{h_1}= 0.5, \frac{a}{h_1}=0.25,\frac{b}{h_1}=0.25,\theta =0^{\circ}$]{\label{fig:vb}\includegraphics*[width=0.49\textwidth]{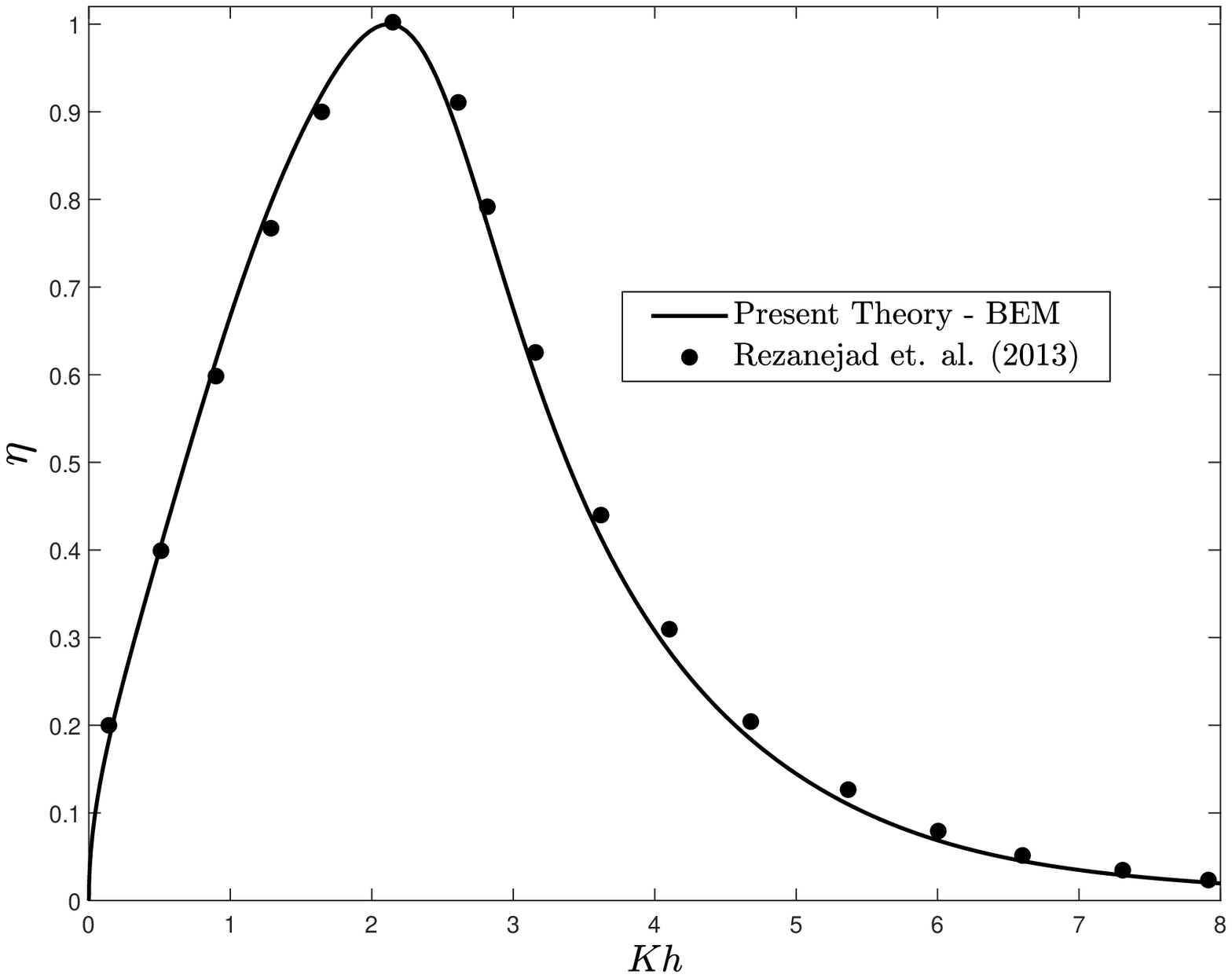}}
	\end{center}\vspace{-0.5cm}
	\caption{Comparison of results using BEM with (a) \citet{evans1995hydrodynamic} and \citet{rezanejad2013stepped} for a flat rigid sea bottom and (b) \citet{rezanejad2013stepped} for a stepped rigid bottom with step height $h_1/4$.}\label{fv}
\end{figure}

\begin{figure}[h!]
	\begin{center}
		\subfigure[Radiation Susceptance ]{\label{fig:2a}\includegraphics*[width=0.49\textwidth]{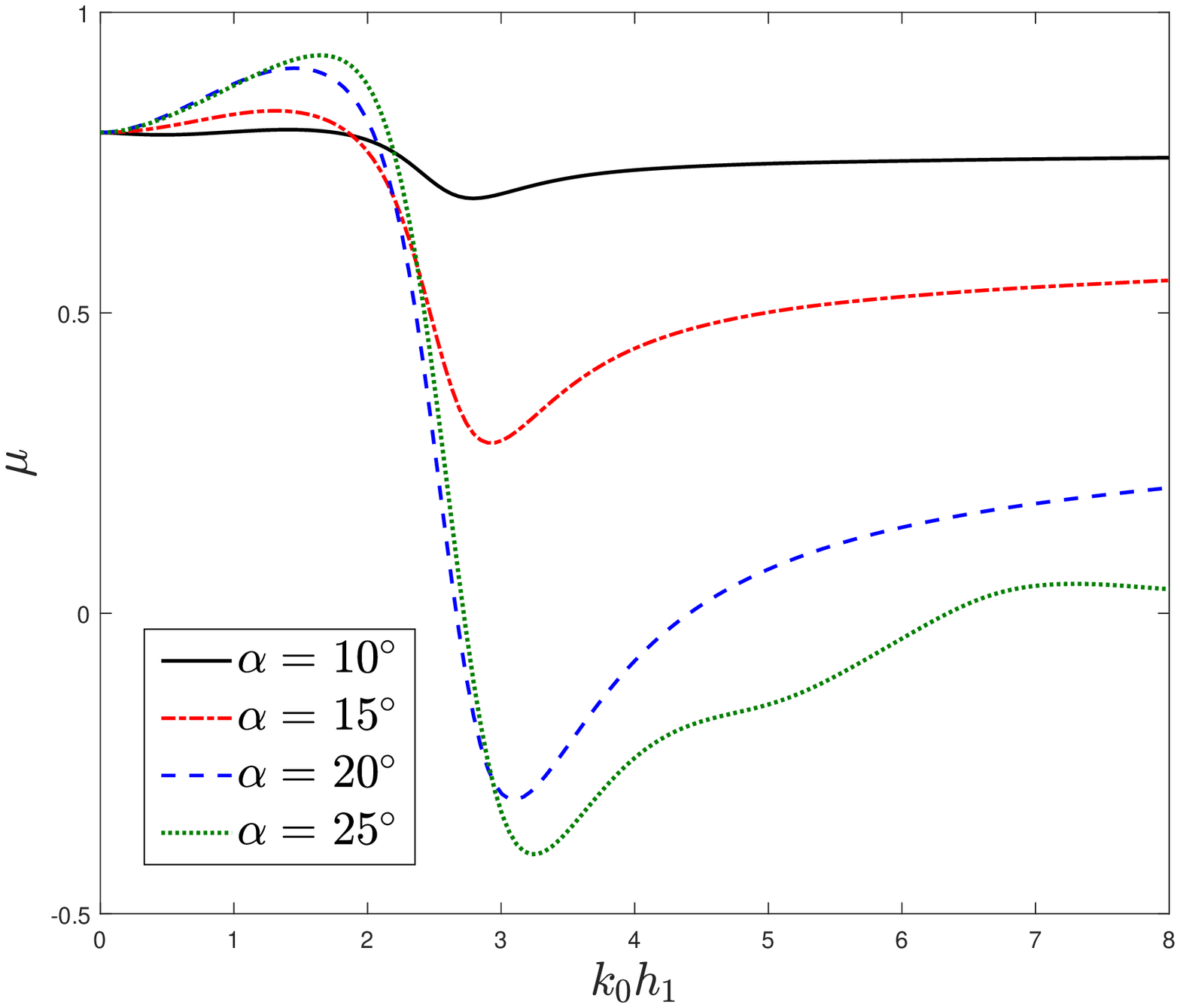}}
		\subfigure[Radiation Conductance]{\label{fig:2b}\includegraphics*[width=0.49\textwidth]{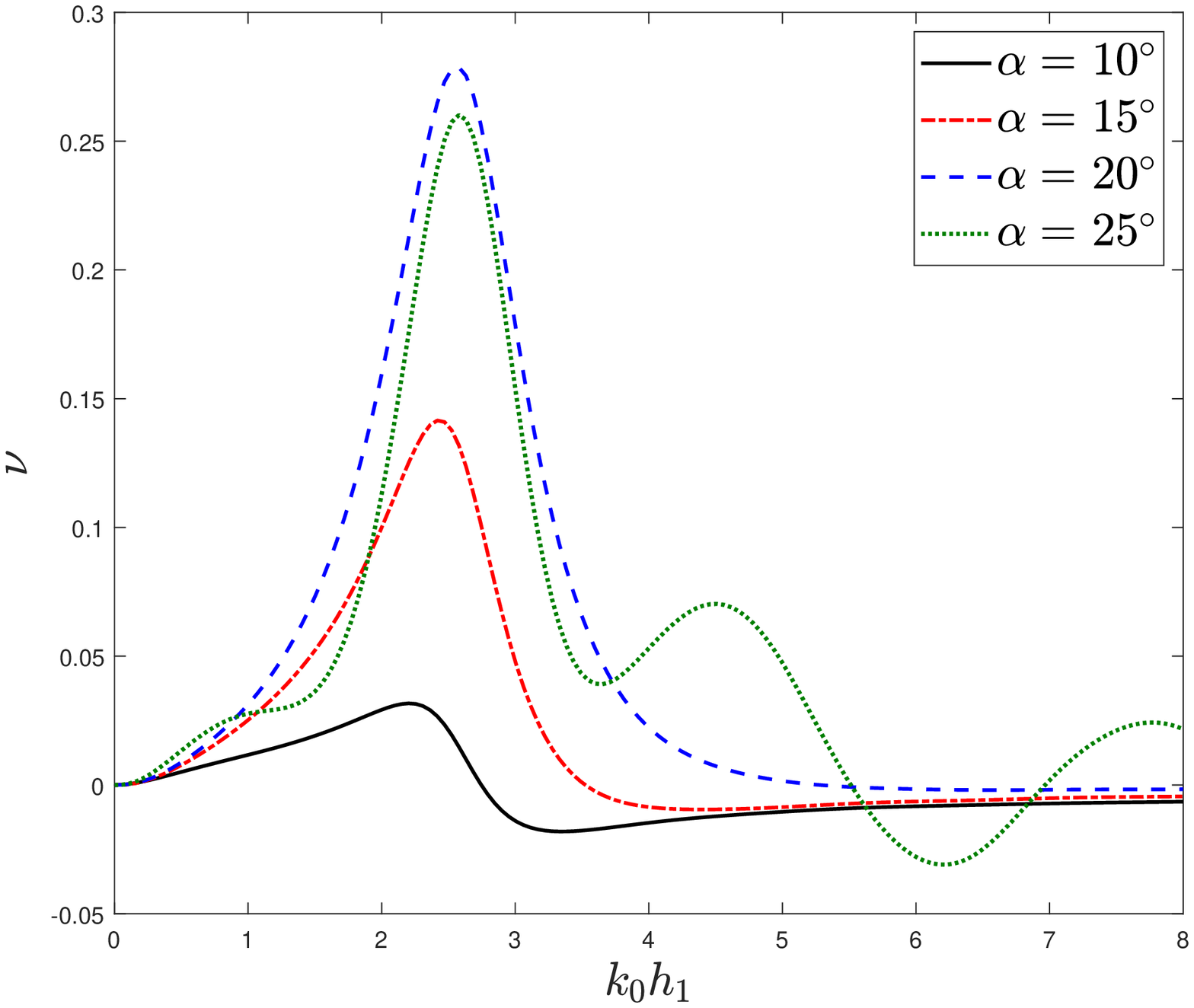}}
	\end{center}\vspace{-0.5cm}
	\caption{Radiation susceptance, conductance, coefficients against non-dimensional wavenumbers with different values of sloping angles for the porous seabed with  $\displaystyle h_1=4,\frac{a}{h_1}=0.25,\frac{b}{h_1}=0.25,\epsilon=0.5,f=0.5,\theta = 10^{\circ}$}\label{f2}
\end{figure}
\begin{figure}[h!]
	\begin{center}		
		\subfigure[$\frac{a}{h_1}=0.25,\frac{b}{h_1}=0.25,\epsilon=0.5,f=0.5,\theta = 10^{\circ}$]{\label{fig:2c}\includegraphics*[width=0.49\textwidth]{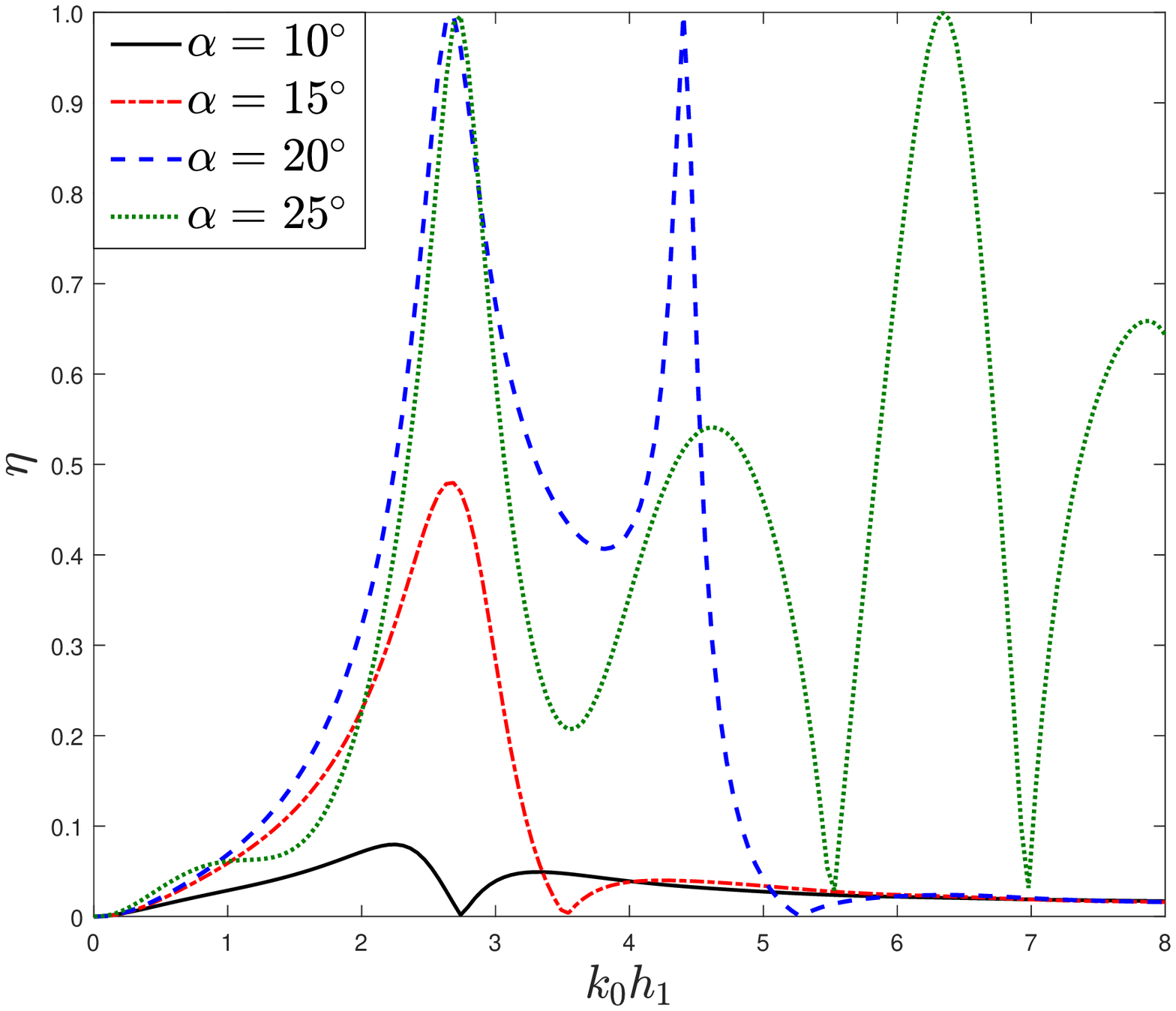}}
		\subfigure[$\alpha=20^{\circ}, \frac{a}{h_1}=0.25,\frac{b}{h_1}=0.25,\epsilon=0.5,f=0.5$]{\label{fig:3c}\includegraphics*[width=0.49\textwidth]{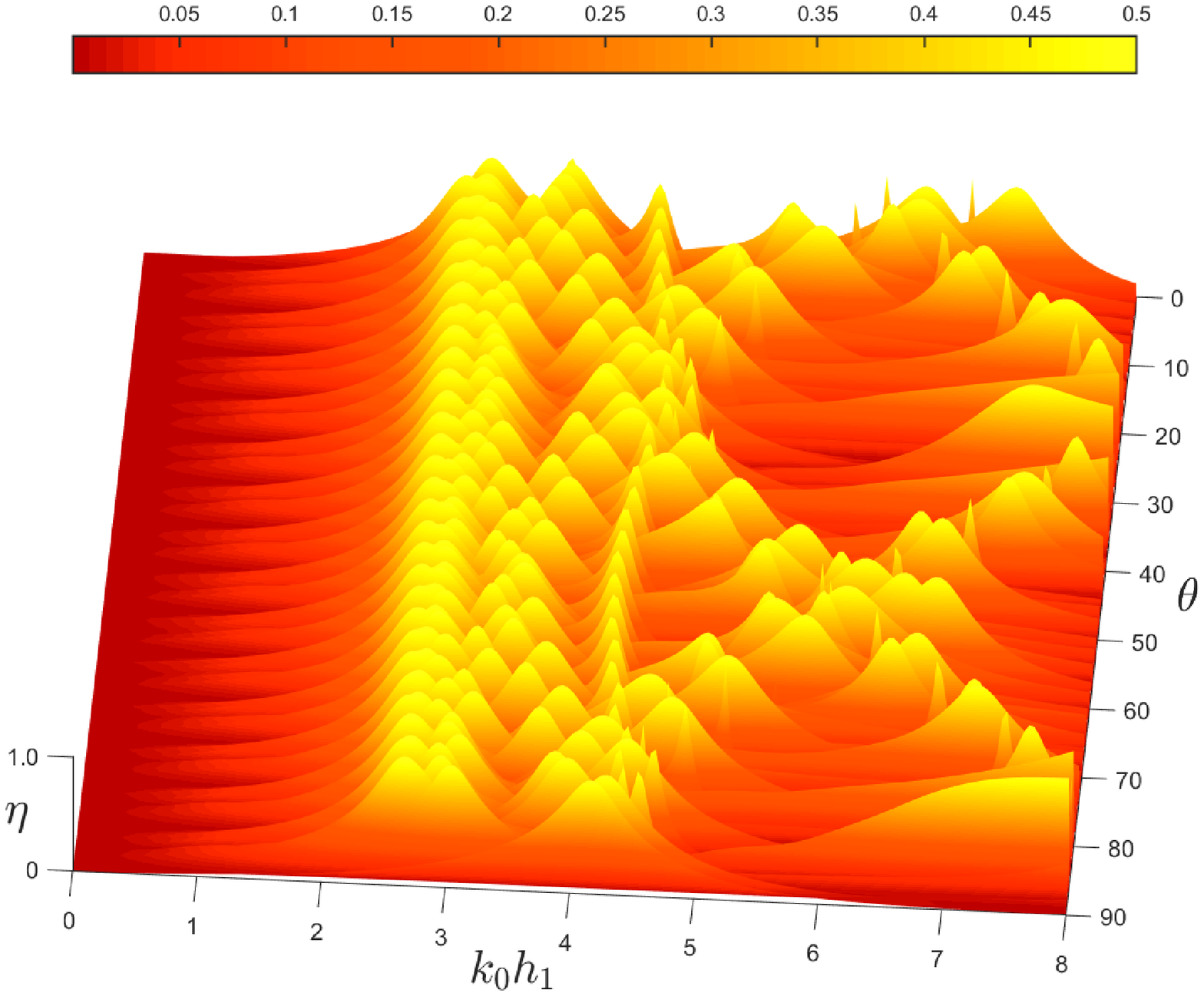}}
	\end{center}\vspace{-0.5cm}
	\caption{Maximum efficiency against non dimensional wavenumbers and incident angle of the wave for (a) different values of sloping angles for the porous seabed, and (b) surface plot  over the whole range of incident wave angles $(k_0h_1,\theta)$. }\label{f3}
\end{figure}

\section{Results and Discussion}
This section describes the preliminary results for regulation of radiation conductance, susceptance and maximum efficiency by a sloping porous bed. The porous bed parameters specifically porosity, frictional coefficient, sloping angle are varied to study their impact on the performance of the OWC. Certain, physical parameters of the system are kept fixed throughout the simulation as $\rho =1025$kg/m$^3$, $h_1=4$m, $g=9.81$m/s$^2$, unless stated otherwise. In particular, the depth of water region at the rigid wall changes due to variation in sloping angle, which is given by $h_2 = h_1 -(L+b)\tan\theta$. Using formulas in Eq. \eqref{procl}, the maximum efficiency of the water column $\eta$ is written as
\begin{equation}
	\eta =\frac{2}{\bigg(1+\big(\frac{\mu}{\nu}\big)^2\bigg)^{\frac{1}{2}}+1}.\label{eff}
\end{equation}
In comparison with a rigid body system, radiation susceptance $\mu$ and conductance $\nu$ are analogous to added mass and radiation damping, respectively, with $\nu$ responsible for energy transfer into the system while $\mu$ contributes to the energy that is not utilized. Eq. \eqref{eff} shows that increase in the system efficiency is necessitated by relative magnitudes of high radiation conductance and low radiation susceptance.

\begin{figure}[h!]
	\begin{center}		
		\subfigure[Radiation susceptance]{\label{fig:}\includegraphics*[width=0.49\textwidth]{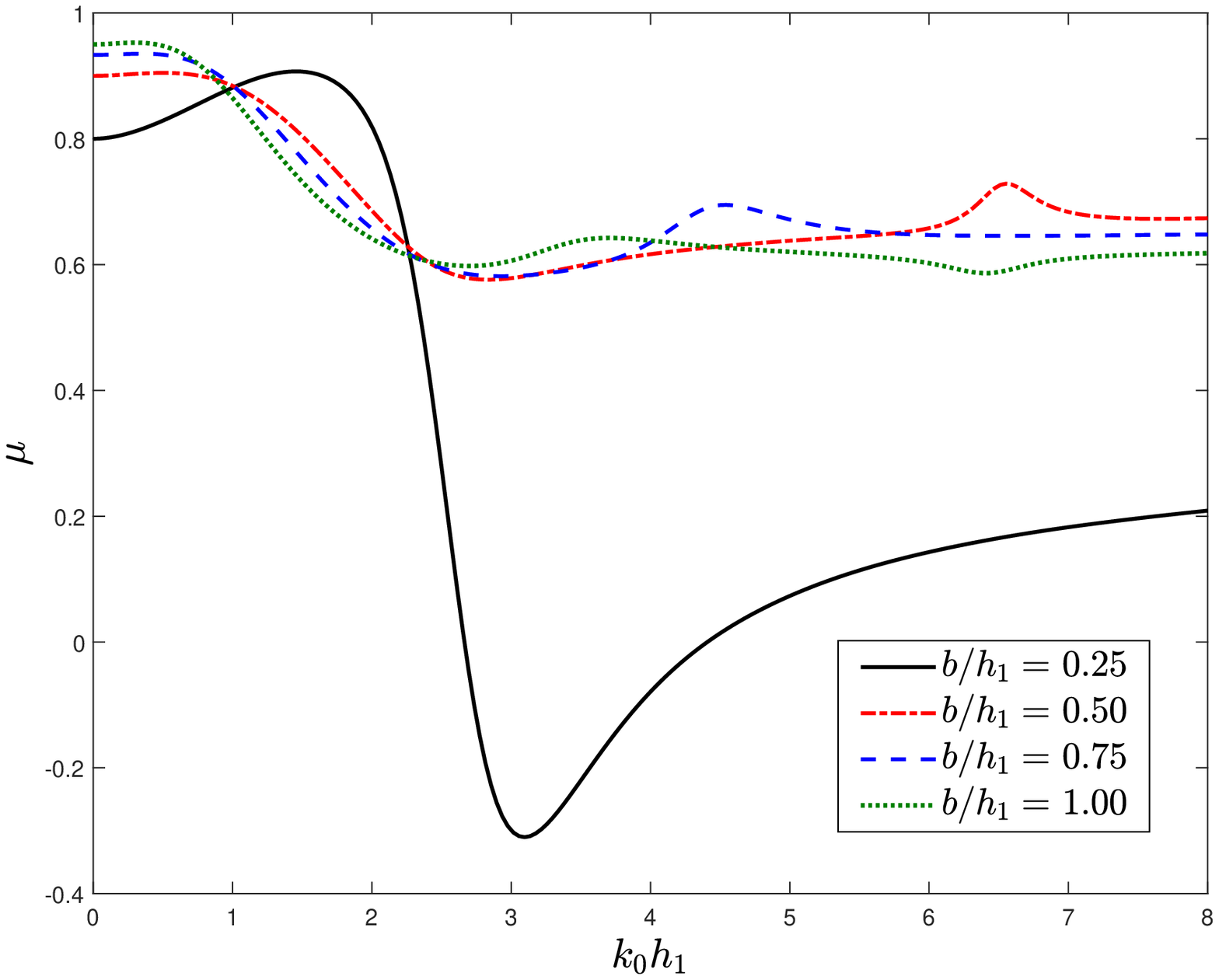}}
		\subfigure[Radiation conductance]{\label{fig:}\includegraphics*[width=0.49\textwidth]{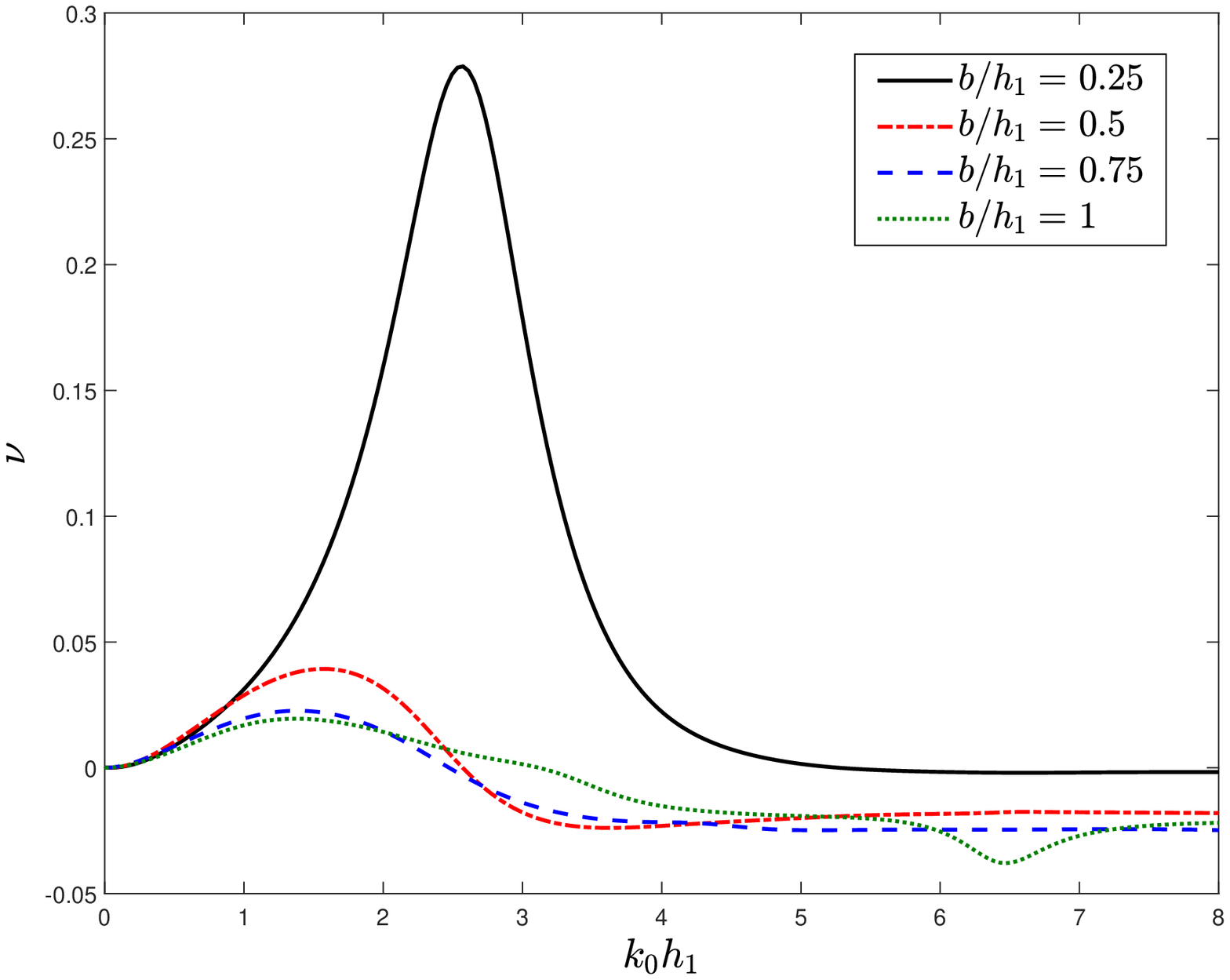}}
		\subfigure[Maximum efficiency]{\label{fig:}\includegraphics*[width=0.49\textwidth]{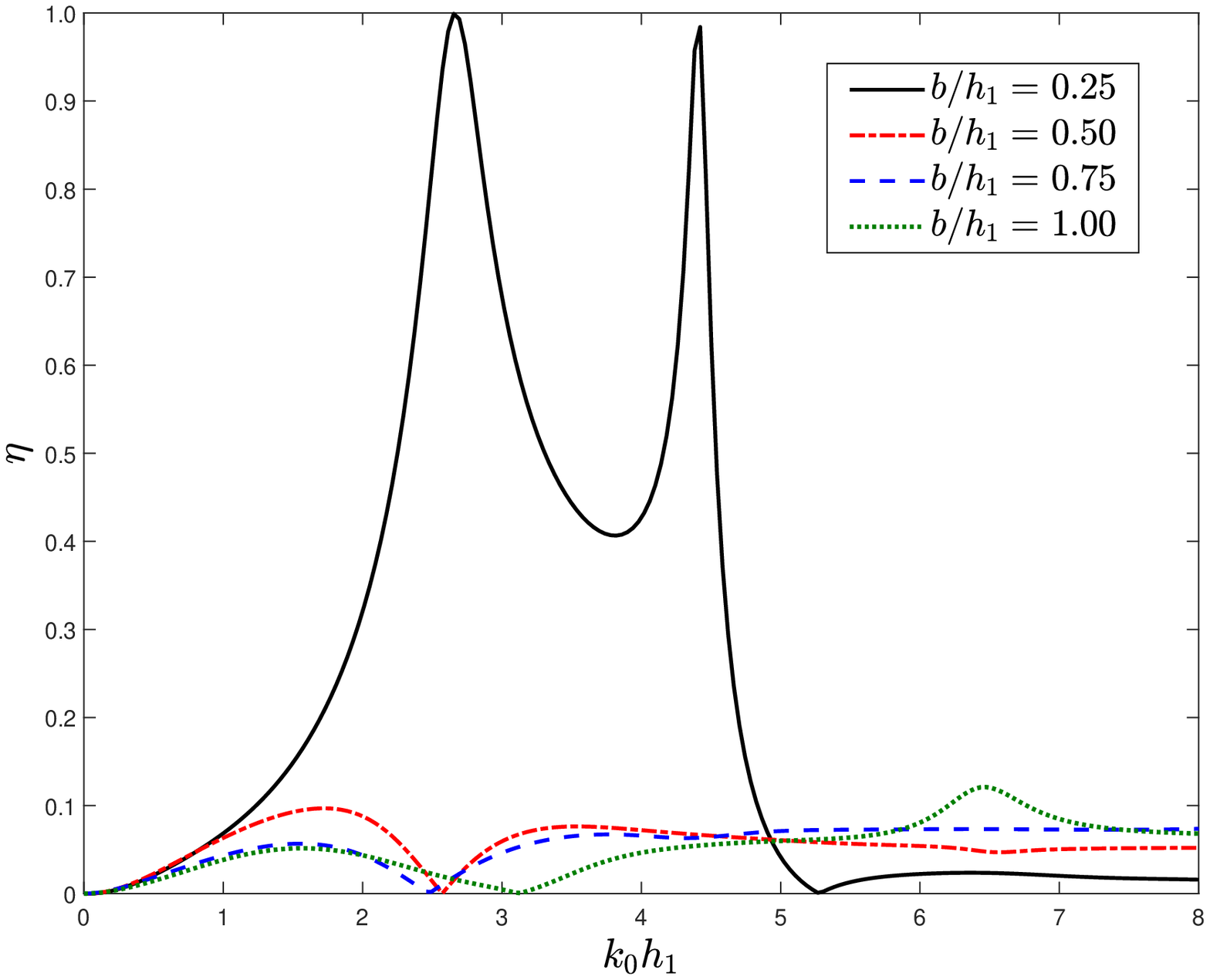}}
	\end{center}\vspace{-0.5cm}
	\caption{Radiation susceptance, conductance, and maximum efficiency coefficients against non-dimensional wavenumbers with different values of wall spacing for the OWC device on the porous seabed with  $\displaystyle\alpha=20^{\circ},\frac{a}{h_1}=0.25,\epsilon = 0.5,f=0.5, \theta = 10^{\circ}$ }\label{f6}
\end{figure}
\begin{figure}[h!] 
	\begin{center}		
		\subfigure[Radiation susceptance]{\label{fig:}\includegraphics*[width=0.49\textwidth]{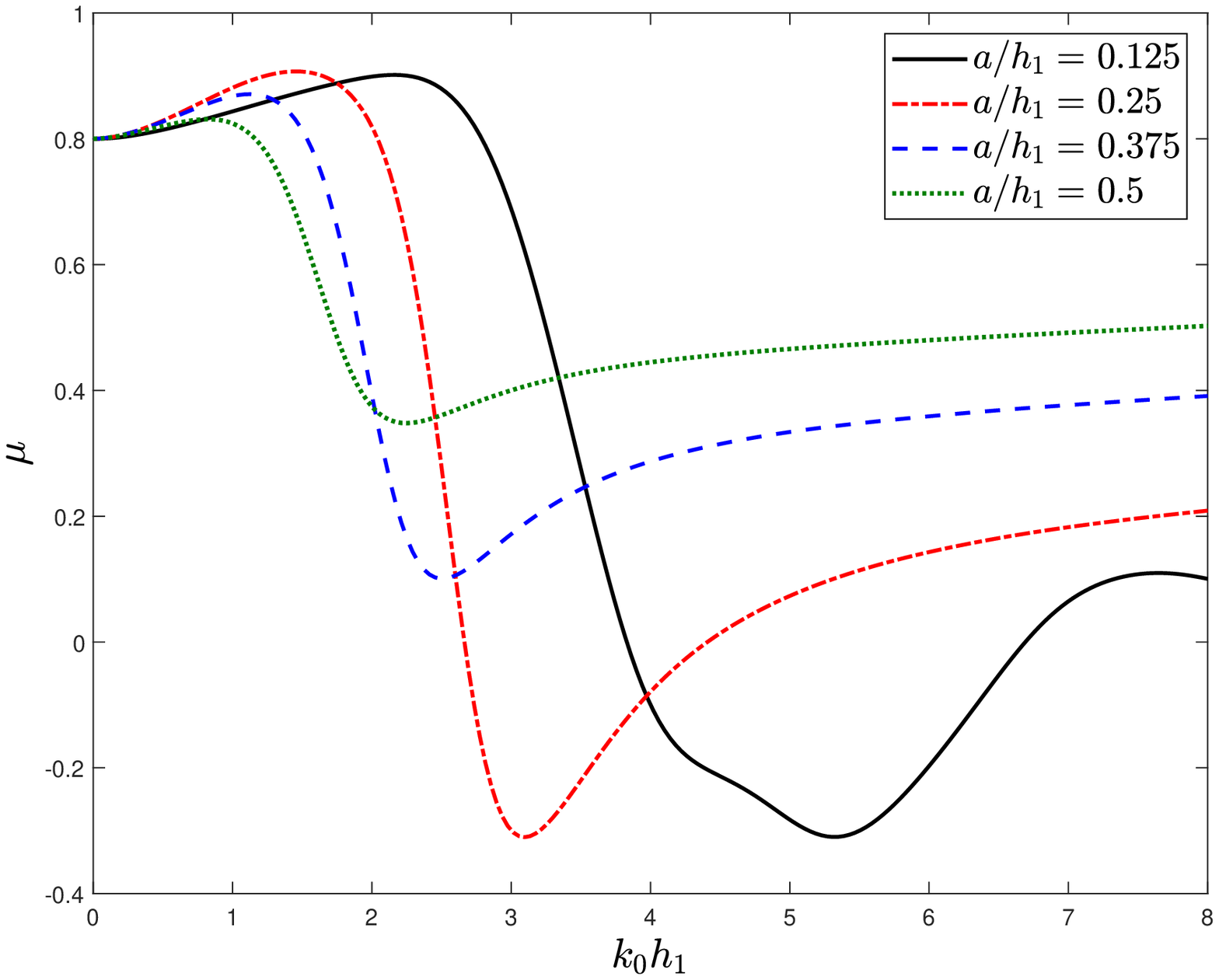}}
		\subfigure[Radiation conductance]{\label{fig:}\includegraphics*[width=0.49\textwidth]{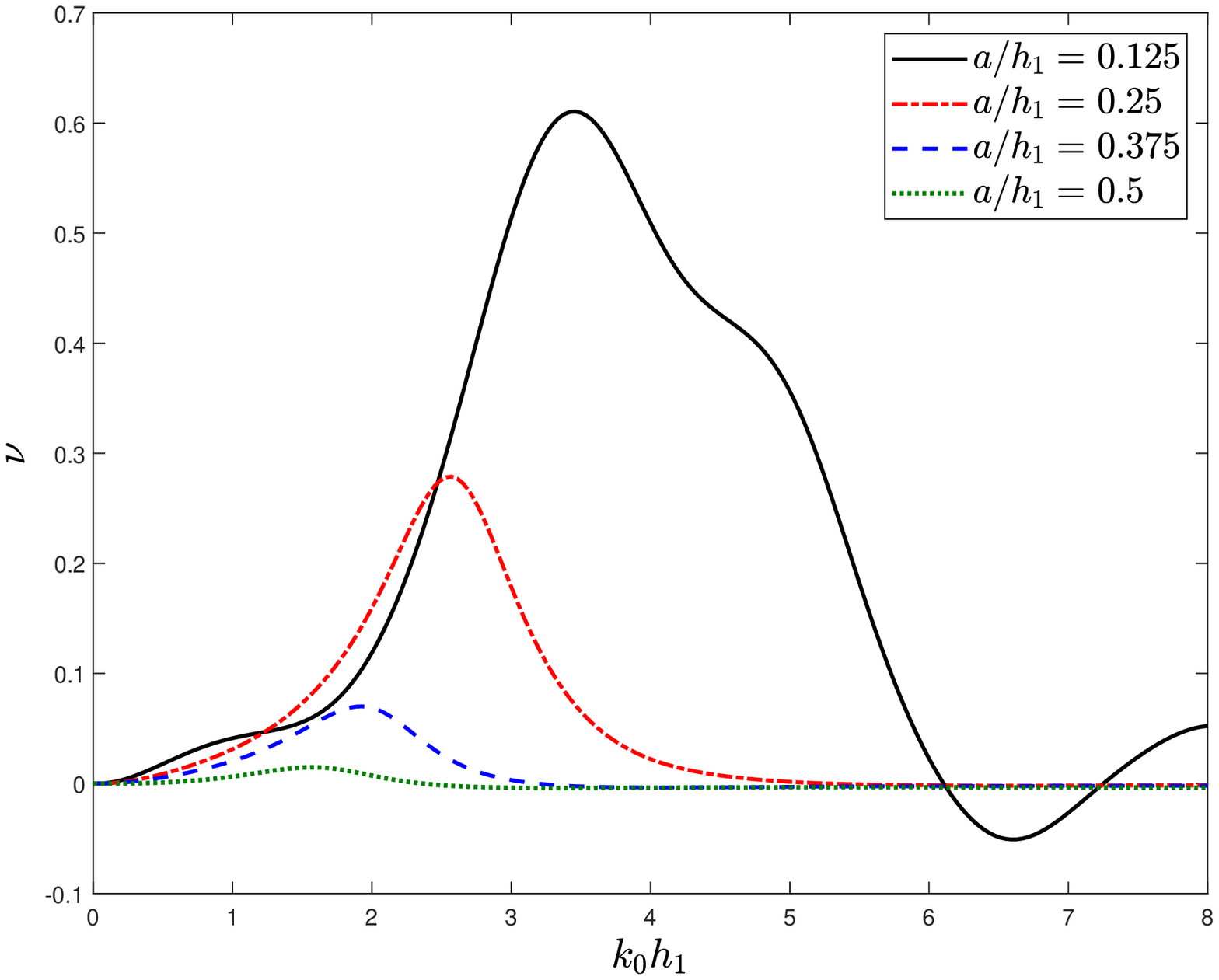}}
		\subfigure[Maximum efficiency]{\label{fig:}\includegraphics*[width=0.49\textwidth]{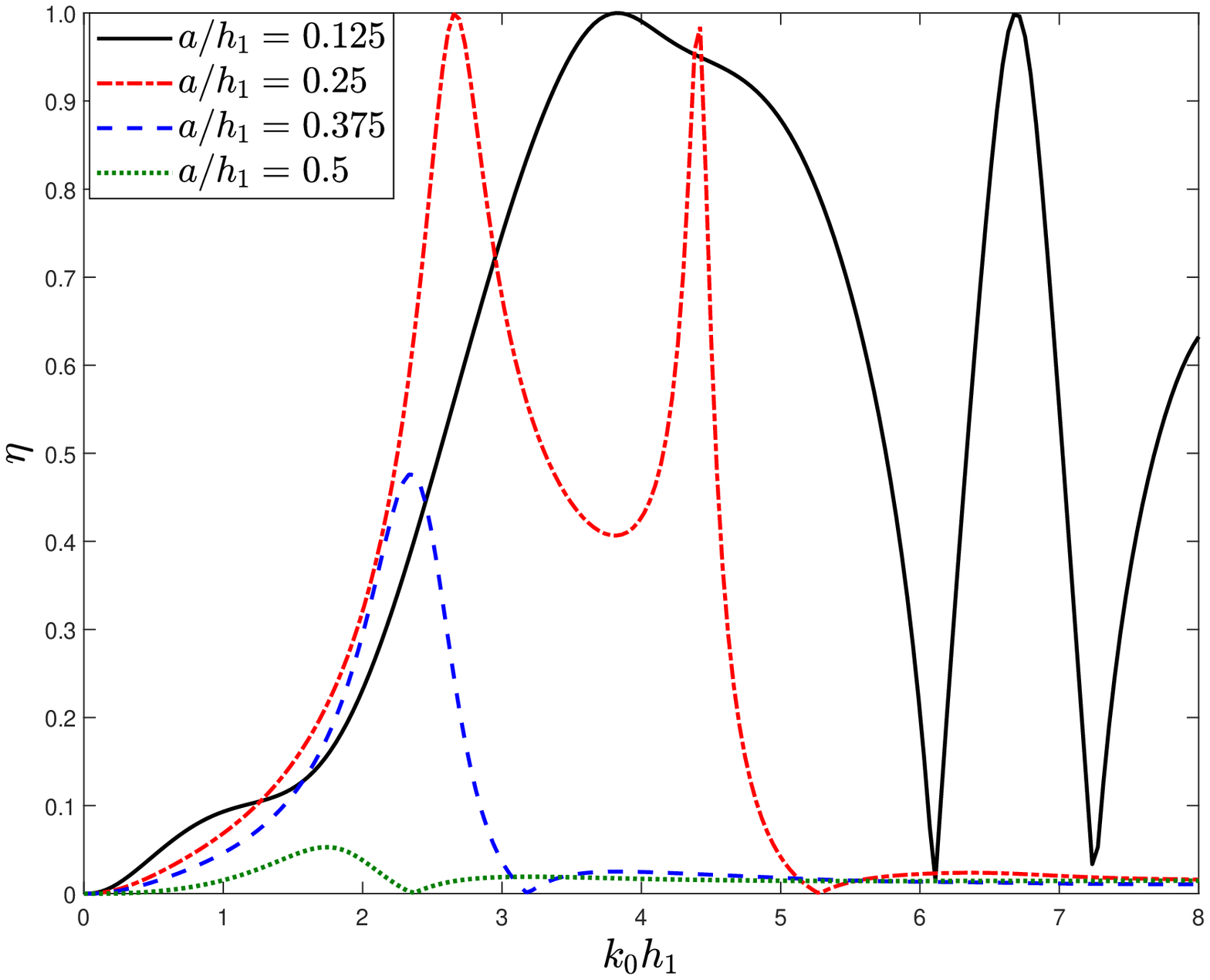}}
	\end{center}
	\caption{Radiation susceptance, conductance, and maximum efficiency coefficients against non-dimensional wavenumbers with different values of depth of the OWC device on the porous seabed with  $\displaystyle\alpha=20^{\circ},b/h_1=0.25,\epsilon = 0.5,f=0.5, \theta = 10^{\circ}$ }\label{f7}
\end{figure}

\begin{figure}[h!]
	\begin{center}		
		\subfigure[Radiation susceptance]{\label{fig:7a}\includegraphics*[width=0.49\textwidth]{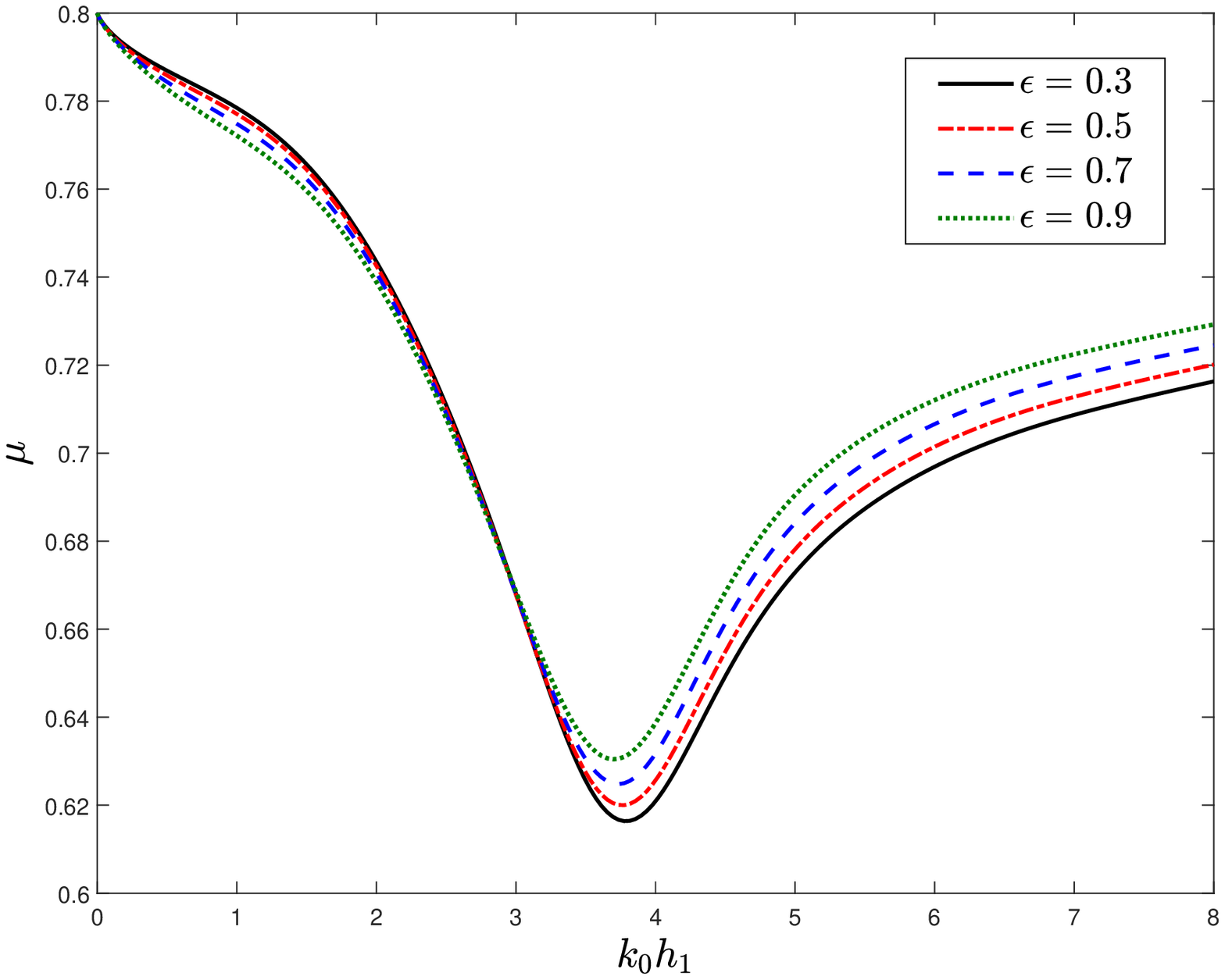}}
		\subfigure[Radiation conductance]{\label{fig:7b}\includegraphics*[width=0.49\textwidth]{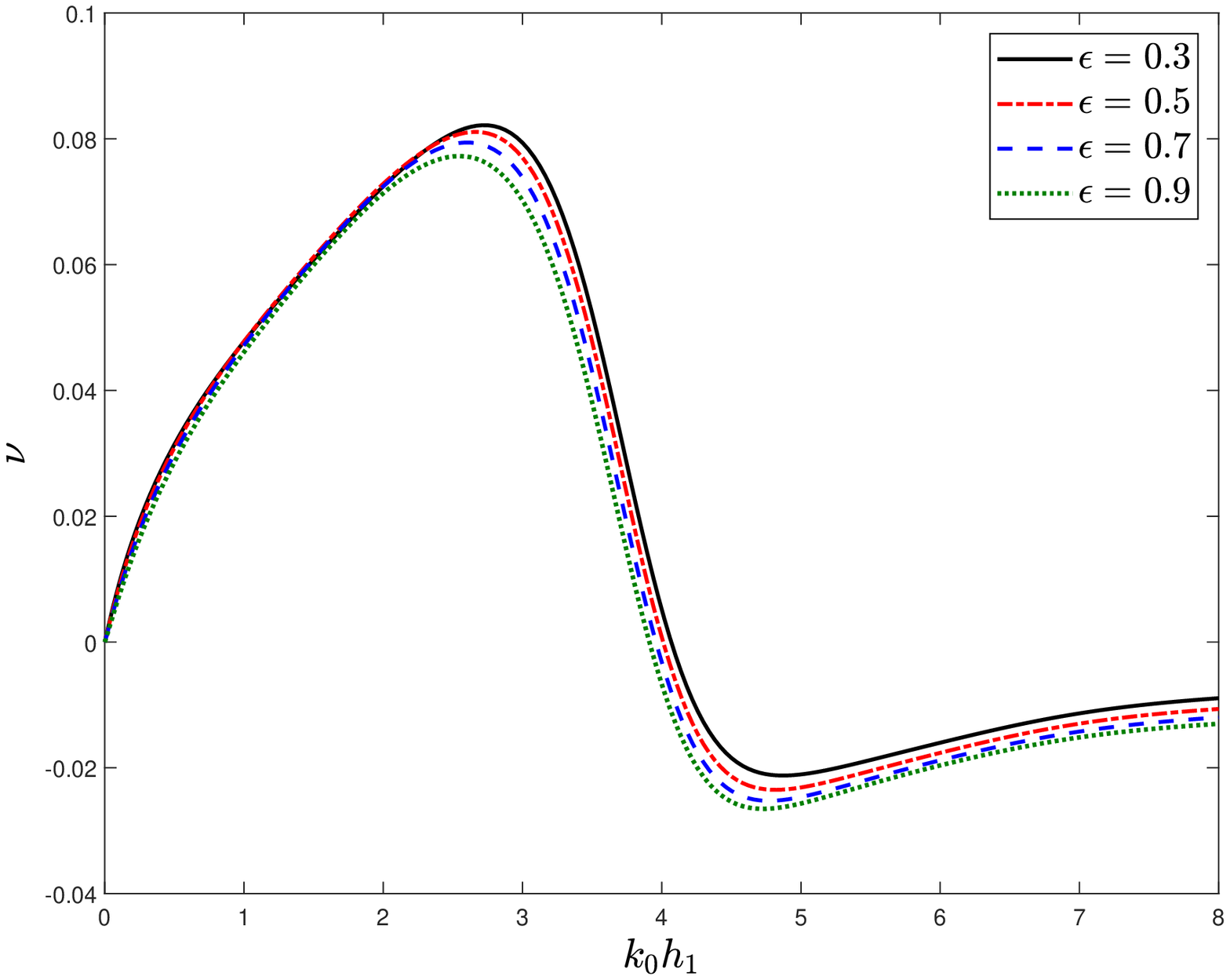}}
		\subfigure[Maximum efficiency]{\label{fig:7c}\includegraphics*[width=0.49\textwidth]{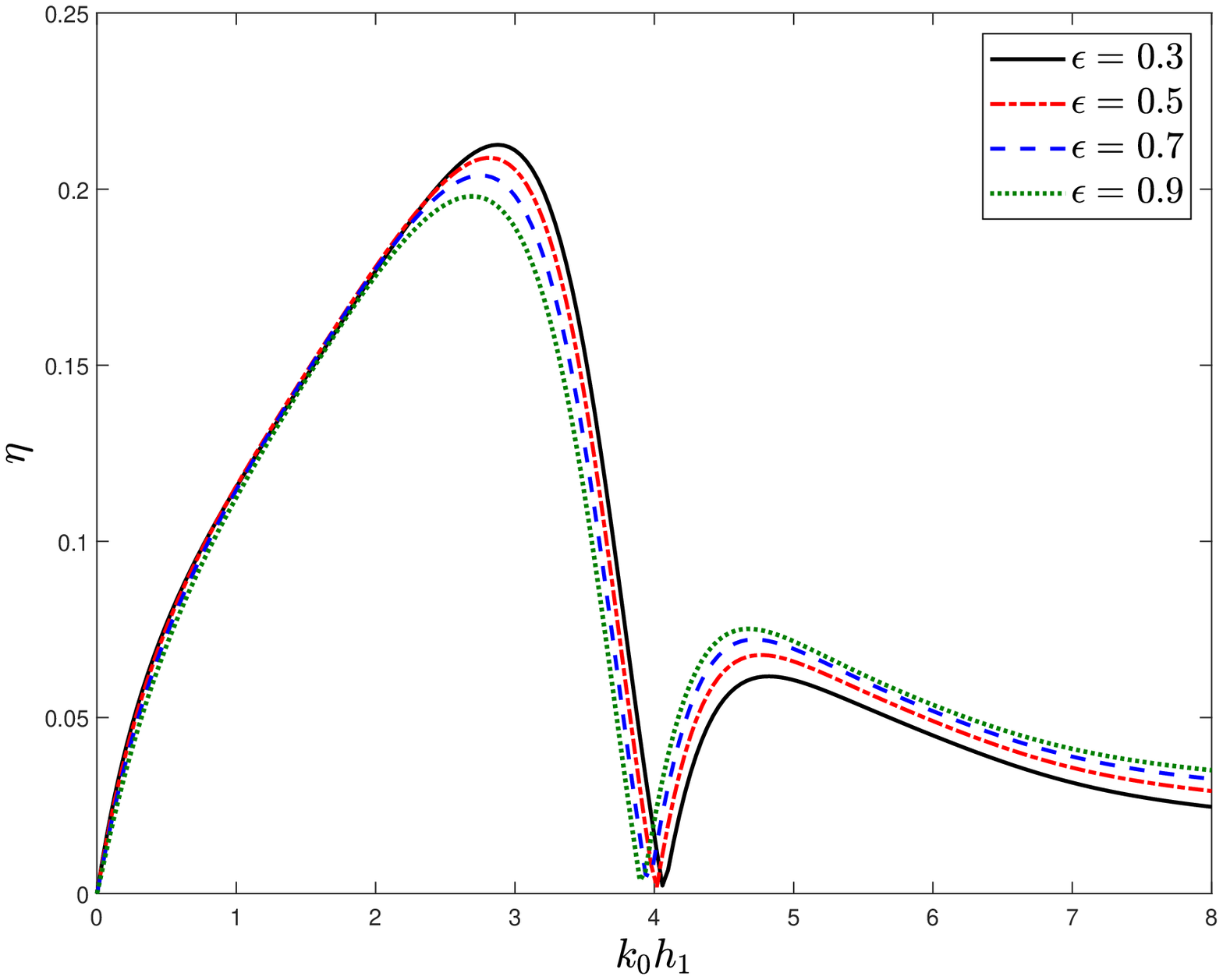}}
	\end{center}\vspace{-0.5cm}
	\caption{Radiation susceptance, conductance, and maximum efficiency coefficients against non-dimensional wavenumbers with different values of porosity for the porous seabed with   $\displaystyle\alpha=10^{\circ},a/h_1=0.125,b/h_1=0.25,f=1.5,\theta = 10^{\circ}$ }\label{f4}
\end{figure}
\begin{figure}[h!]
	\begin{center}		
		\subfigure[Radiation susceptance]{\label{fig:8a}\includegraphics*[width=0.49\textwidth]{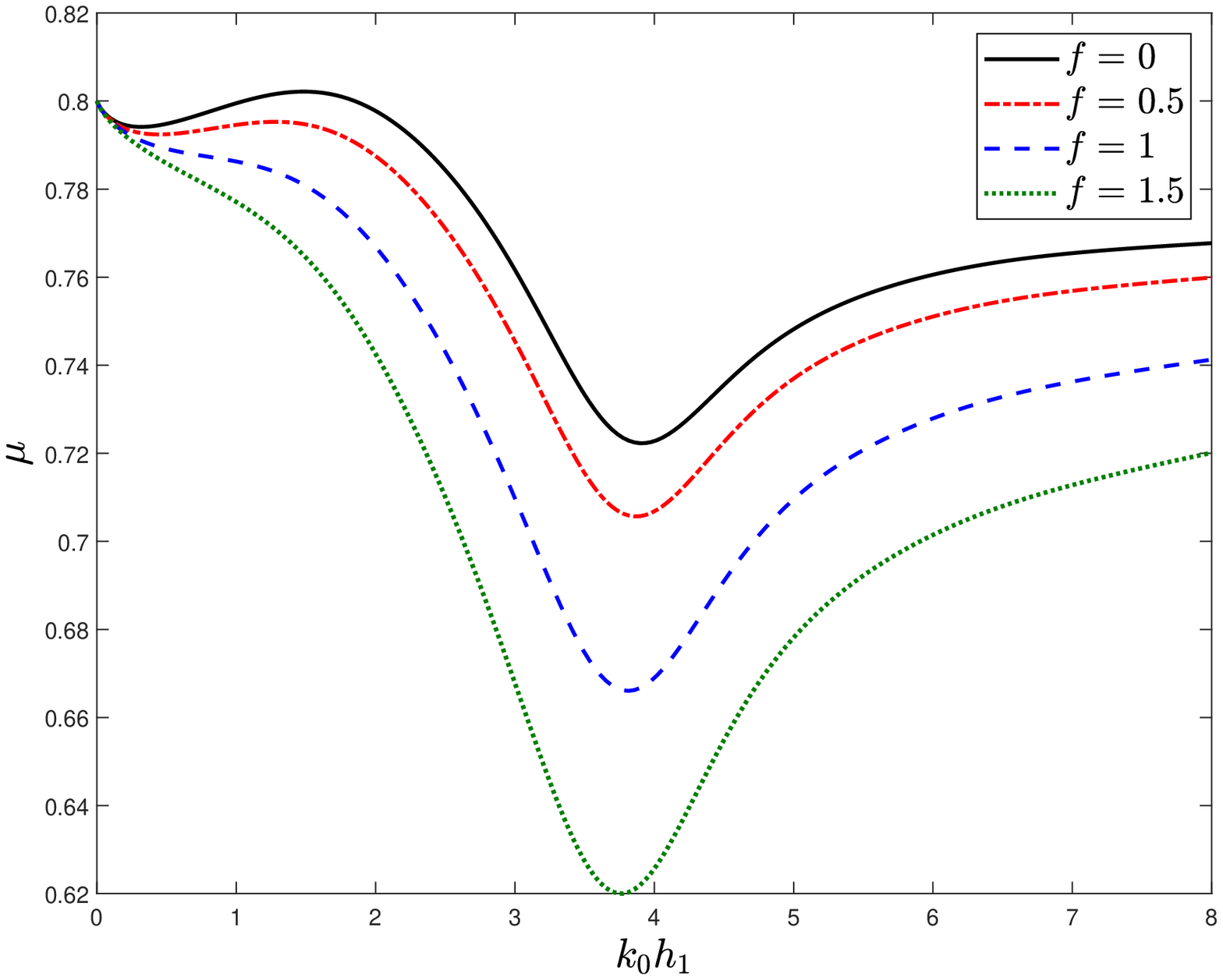}}
		\subfigure[Radiation conductance]{\label{fig:8b}\includegraphics*[width=0.49\textwidth]{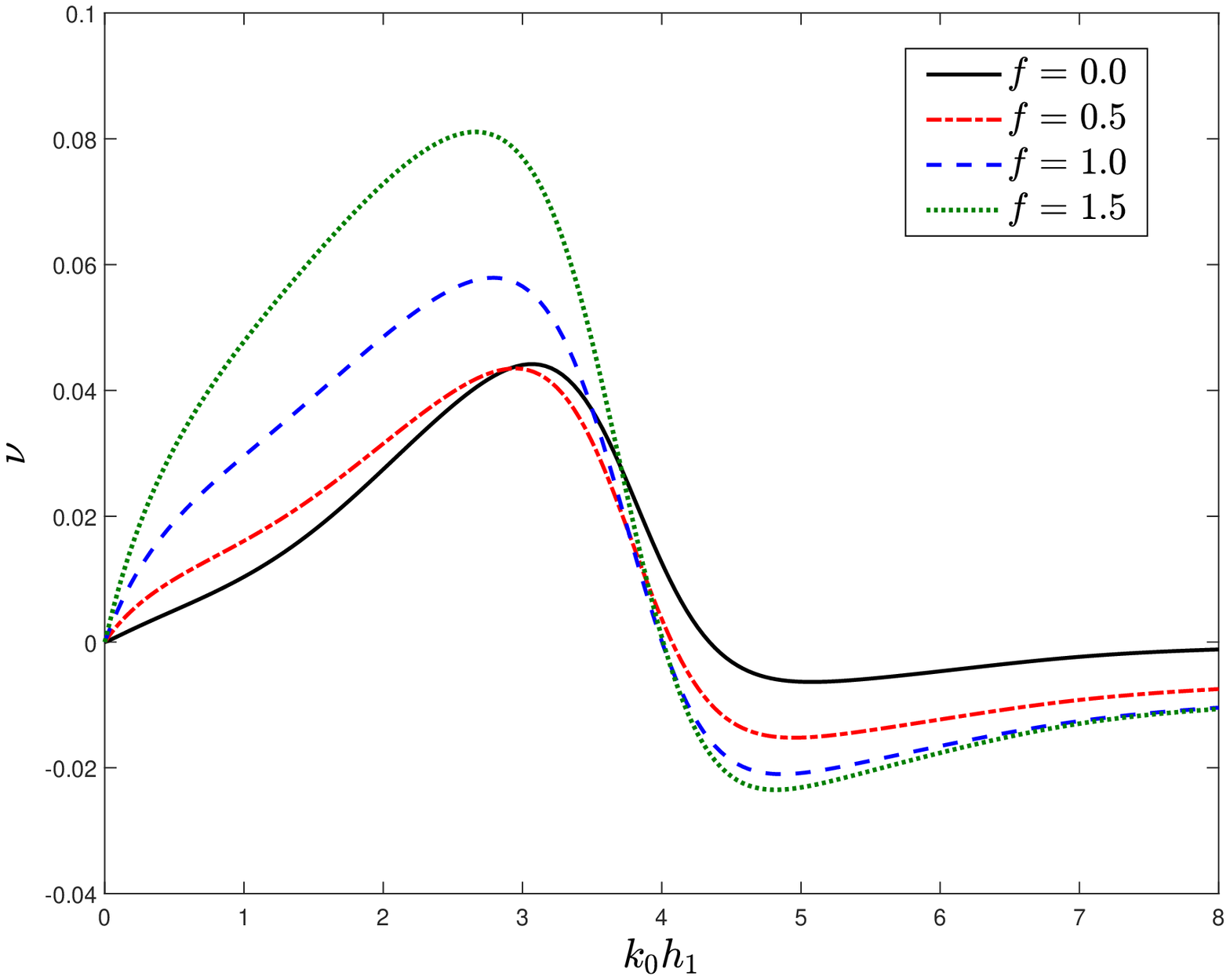}}
		\subfigure[Maximum efficiency]{\label{fig:8c}\includegraphics*[width=0.49\textwidth]{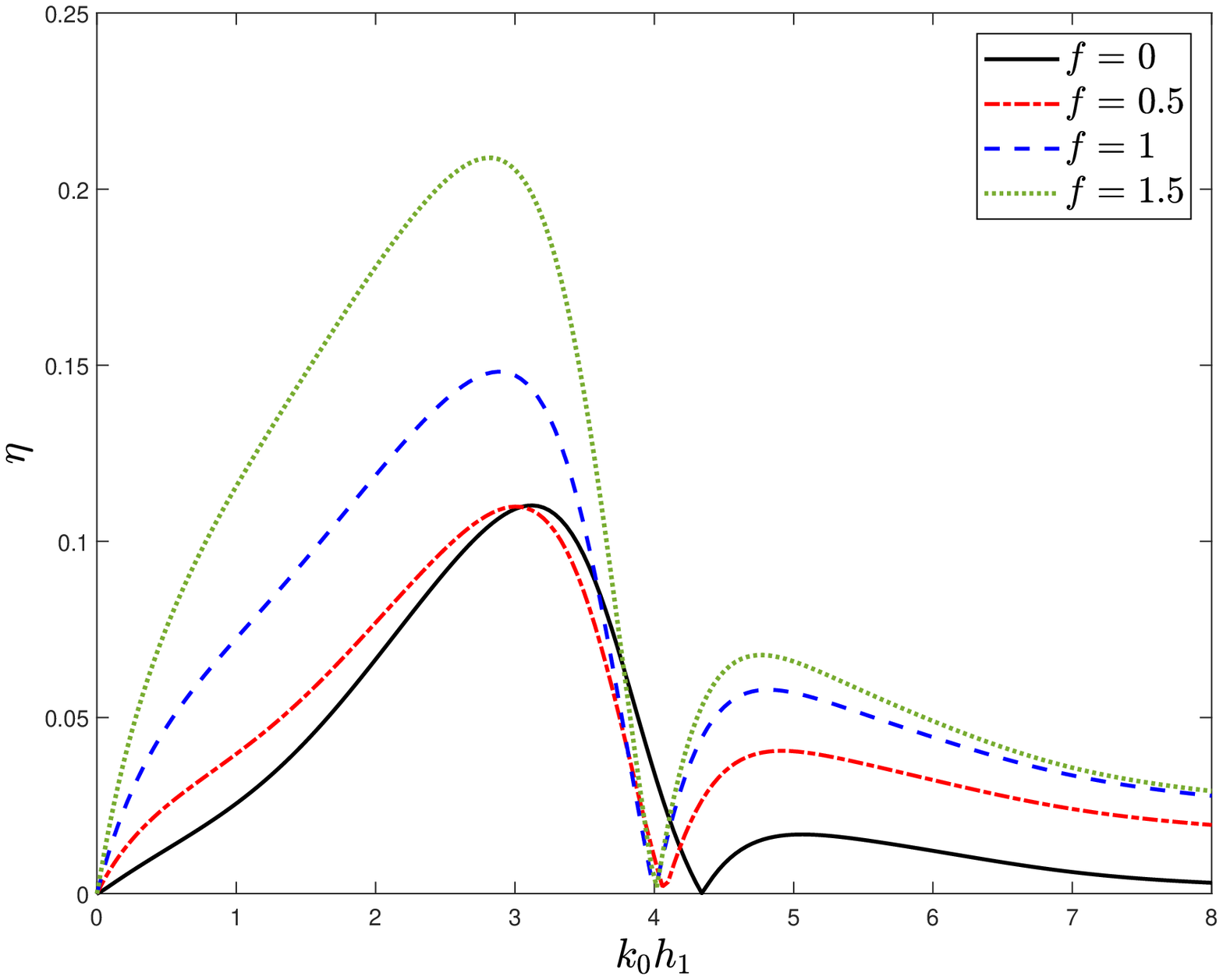}}
	\end{center}\vspace{-0.5cm}
	\caption{Radiation susceptance, conductance, and maximum efficiency coefficients against non-dimensional wavenumbers with different values of friction for the porous seabed with  $\displaystyle\alpha=10^{\circ},a/h_1=0.125,b/h_1=0.25,f=1.5,\epsilon=0.5,\theta = 10^{\circ}$}\label{f5}
\end{figure}

In Fig. \ref{fv}, the results for the maximum efficiency $\eta$ against $Kh$, where $K = \omega^2/g, h = h_1$,  are compared with the results in special case of a flat sea bottom without a porous sloping bed (Fig. \ref{fig:va}), calculated analytically using Galerkin technique in \citet{evans1995hydrodynamic} and numerically using BEM in \citet{rezanejad2013stepped}. In  Fig. \ref{fig:vb}, the results are compared for a stepped sea bottom with step height $h_1/4$ and the OWC device at a distance $b$ from the step edge, calculated using BEM, as in \citet{rezanejad2013stepped}. The comparisons performed depict these results in good agreement, thereby establishing the validity of our method.

Figs. \ref{fig:2a} and \ref{fig:2b} depict $\mu$ and $\nu$ against non-dimensional wavenumbers for different values of sloping angles for the seabed. The resonant frequency of peak efficiency corresponds to the frequency for maxima of conductance $\nu$ and zeros of the susceptance $\mu$. In general, the magnitude of susceptance is minimized near dimensionless frequency $k_0h_1\sim3$ while that of conductance is maximized between frequencies $2\leq k_0h_1\leq3$. On increasing the slope, the optima are further extremized in both cases, however, oscillatory behaviour arises in $\mu$ and $\nu$ for higher slopes of the porous seabed.

The effect of slope on efficiency is more profound (Fig. \ref{fig:2c}), as a $10^{\circ}$ increase in the slope increases the maximum efficiency of the system 100 times ($\alpha = 10^{\circ}\rightarrow 20^{\circ}  $).  The higher slope angle also results in multiple resonant frequencies for the OWC device, and a broader range of frequencies at which the efficiency is relatively higher. Generally, the resonant frequency decreases when water depth is decreased inside the OWC (see \citet{rezanejad2013stepped}), in this case however, the resonant frequency is independent of water depth corresponding to increase in slope of the porous bed, which decreases the average depth water column inside the OWC. Physically, this may be due to porous nature of the seabed because motion of the water particles does not change to completely elliptical as would be the case for a rigid slope, thereby not requiring a longer travel path in one period of motion. Finally, Fig. \ref{fig:3c} depicts the variation in OWC efficiency over the full wave spectrum and for incident wave angles $0^{\circ}\leq\theta\leq90^{\circ}$. In general, for oblique waves, OWC with a sloping porous seabed has high efficiency for frequencies $2\leq k_0h_1\leq4$ with multiple resonant frequencies in between for various oblique waves. However, for shortwaves with frequencies $k_0h_1\geq4$, the OWC efficiency is relatively low  with one or two resonant frequencies depending on the oblique wave angle.

Panels \ref{f6}(a), \ref{f6}(b) and \ref{f6}(c) show the radiation suceptance, conductnce and maximum efficiency coefficients, respectively, against non-dimensional wavenumber for various non-dimensional wall spacings for the OWC device. It is observed that for $b/h_1$ = 1/4, maximum efficiency occurs at the wavenumbers corresponding with zeros of $\mu$ and maxima of $\nu$. However, upon increasing width of the wall spacing, the efficiency of OWC device over a sloping porous seabed reduces significantly. This may be due to dissipation of the wave energy by the porous seabed, which reduces the amaount of wave energy available for conversion. Additionally, maximum efficiency reduces to zero corresponding to zeros of $\nu$.

Figs. \ref{f7}(a), \ref{f7}(b) and \ref{f7}(c) show the radiation suceptance, conductance and maximum efficiency coefficients, respectively, against non-dimensional wavenumber for various non-dimensional barrier depth for the OWC device over a sloping porous seabed. In particular, for the slope $\displaystyle\alpha = 10^{\circ}, a/h_1 = 0.125,0.25$, maximum efficiency reaches peak values of $\eta=1$ at multiple wavenumbers. Moreover for OWC with $a/h=0.125$, waves over a large range of wavenumbers are efficiently used for energy conversion.
As the barrier depth of the OWC device is increased the efficiency decreases significantly. This is due to the fact that waves with sufficient energy to drive the OWC mechanism are concentrated near the free surface and larger barriers hinder the amount of energy entering the oscillating water column, while some energy is also dissipated by the sloping porous bed.

Other defining structural parameters of the sloping porous seabed are structural porosity and frictional coefficient, the variational effects of which are illustrated in Figs. \ref{f4} and \ref{f5} for different base structural parameters. Porosity of the seabed has opposite effects on efficiency of the OWC for long and shortwaves. While higher porosity results in less efficient OWC when faced with long waves, the efficiency increases for short waves (Fig. \ref{fig:7c}). In particular, for the configuration $\displaystyle\alpha=10^{\circ},a/h_1=0.125,b/h_1=0.25,f=1.5,\theta = 10^{\circ}$, the maximum efficiency is reduced to zero near $k_0h =4$. This also corresponds to the zero of radiation conductance for this configuration (Fig. \ref{fig:7a}).  The reason is that an increase in porosity results in more dissipation of wave energy by the porous seabed, thereby reducing the energy available to OWC device for conversion.

Fig. \ref{f5} depicts variational effects of frictional coefficient of the sloping porous seabed on $\mu, \nu$ and $\eta$ in the same configuration as in Fig. \ref{f4} with $\displaystyle\alpha=10^{\circ},a/h_1=0.125,b/h_1=0.25,\epsilon=0.5,\theta = 10^{\circ}$. However, contrary to porosity effects, the increase in value of the frictional coefficient favors the efficiency of the OWC over almost whole of the wavenumber spectrum. Similar to Fig. \ref{fig:7c}, the maximum efficiency is reduced to zero  near $k_0h =4$ in Fig. \ref{fig:8c}, corresponding to zeros of radiation conductance in Fig. \ref{fig:8b}, even when the corresponding radiation susceptance is minimized as in \ref{fig:8a}. Thus, changes in friction and porosity of the sloping seabed have opposing effects on maximum efficiency of the OWC for long waves and similar effects on $\eta$ for short waves.

\section{Conclusion}
This paper describes the impact of a sloping porous bottom on an oscillating water column device set up near a vertical wall, with oblique incident waves. The device is approximated by a vertical barrier near a rigid wall and placed over a sloping porous structure, while the boundary value problem is tackled using multi-domain BEM. Coefficients of radiation susceptance, conductance, and maximum efficiency are evaluated and compared for various configurations. The resonant frequency of peak efficiency corresponds to the frequency for maxima of conductance $\nu$ and zeros of the susceptance $\mu$. Additionally, maximum efficiency reduces to zero corresponding to zeros of $\nu$.

In general, the magnitude of susceptance is minimized near dimensionless frequency $2\leq k_0h_1\leq4$ for various configurations while that of conductance is maximized in a similar and often overlapping wavenumber range. The effect of slope on efficiency is more profound (Fig. \ref{fig:2c}), as a $10^{\circ}$ increase in the slope increases the maximum efficiency of the system 100 times ($\alpha = 10^{\circ}\rightarrow 20^{\circ}  $).  The higher slope angle also results in multiple resonant frequencies for the OWC device, and a broader range of frequencies at which the efficiency is relatively higher. Moreover, for oblique waves, OWC device with a sloping porous seabed has high efficiency for frequencies $2\leq k_0h_1\leq4$ with multiple resonant frequencies in between, for various oblique waves. However, for shortwaves with frequencies $k_0h_1\geq4$, the OWC device efficiency is relatively low  with one or two resonant frequencies depending on the oblique wave angle. Upon increasing width of the wall spacing, the efficiency of OWC device over a sloping porous seabed reduces significantly while increasing barrier depth of the OWC device decreases the peak efficiency. 

Additionally,  the changes in frictional coefficient and porosity of the sloping porous seabed have opposing effects on maximum efficiency of the OWC for long waves and congruent effects on $\eta$ for short waves. Increasing the porosity of the seabed has decrease the efficiency of the OWC for long waves while increasing that for short waves. However, contrary to porosity effects, the increase in value of the frictional coefficient favors the efficiency of the OWC over almost whole of the wavenumber spectrum. Furthermore, its is evident that presence of a porous sloping seabed dissipates a significant amount of incoming wave-energy, which is parametrized by porosity and frictional coefficient. 

Thus, the porous sloping bed characteristics can be optimized to design and implement an effective OWC device as a combination of a wave-energy converter and a breakwater for protecting the near-shore marine facilities, at a place with the geographical limitations of a porous sloping sea floor and an array of wave conditions. 

\section*{Acknowledgment}
HB gratefully acknowledges the financial support from SERB, Department of Science and Technology, Government of India through “CRG” project, Award No. CRG/2018/004521. 

\section*{Declaration of interests}

The authors report no conflict of interest.

\textbf{Data Availability:} The data that supports the findings of this study are available within the article, highlighted in each of the figure captions and corresponding discussions. 


\bibliography{OWC_ref}

\begin{thebibliography}{15}
\providecommand{\natexlab}[1]{#1}
\providecommand{\url}[1]{\texttt{#1}}
\expandafter\ifx\csname urlstyle\endcsname\relax
  \providecommand{\doi}[1]{doi: #1}\else
  \providecommand{\doi}{doi: \begingroup \urlstyle{rm}\Url}\fi

\bibitem[Boyle(1996)]{boyle1996renewable}
Godfrey Boyle.
\newblock \emph{Renewable energy: power for a sustainable future}, volume~2.
\newblock Oxford University Press, 1996.

\bibitem[Antonio(2010)]{antonio2010wave}
F~de~O Antonio.
\newblock Wave energy utilization: A review of the technologies.
\newblock \emph{Renewable and sustainable energy reviews}, 14\penalty0
  (3):\penalty0 899--918, 2010.

\bibitem[Heath(2012)]{heath2012review}
TV~Heath.
\newblock A review of oscillating water columns.
\newblock \emph{Phil. Trans. R. Soc. A}, 370\penalty0 (1959):\penalty0
  235--245, 2012.

\bibitem[Evans(1976)]{evans1976theory}
DV~Evans.
\newblock A theory for wave-power absorption by oscillating bodies.
\newblock \emph{Journal of Fluid Mechanics}, 77\penalty0 (1):\penalty0 1--25,
  1976.

\bibitem[Evans(1981)]{evans1981power}
DV~Evans.
\newblock Power from water waves.
\newblock \emph{Annual review of Fluid mechanics}, 13\penalty0 (1):\penalty0
  157--187, 1981.

\bibitem[Mei(1976)]{mei1976power}
Chiang~C Mei.
\newblock Power extraction from water waves.
\newblock \emph{Journal of Ship Research}, 20:\penalty0 63--66, 1976.

\bibitem[Evans(1978)]{evans1978oscillating}
DV~Evans.
\newblock The oscillating water column wave-energy device.
\newblock \emph{IMA Journal of Applied Mathematics}, 22\penalty0 (4):\penalty0
  423--433, 1978.

\bibitem[Evans(1982)]{evans1982wave}
DV~Evans.
\newblock Wave-power absorption by systems of oscillating surface pressure
  distributions.
\newblock \emph{Journal of Fluid Mechanics}, 114:\penalty0 481--499, 1982.

\bibitem[Falnes and McIver(1985)]{falnes1985surface}
J~Falnes and P~McIver.
\newblock Surface wave interactions with systems of oscillating bodies and
  pressure distributions.
\newblock \emph{Applied Ocean Research}, 7\penalty0 (4):\penalty0 225--234,
  1985.

\bibitem[Evans and Porter(1995)]{evans1995hydrodynamic}
DV~Evans and R~Porter.
\newblock Hydrodynamic characteristics of an oscillating water column device.
\newblock \emph{Applied Ocean Research}, 17\penalty0 (3):\penalty0 155--164,
  1995.

\bibitem[Deng et~al.(2013)Deng, Huang, and Law]{deng2013wave}
Zhengzhi Deng, Zhenhua Huang, and Adrian~WK Law.
\newblock Wave power extraction by an axisymmetric oscillating-water-column
  converter supported by a coaxial tube-sector-shaped structure.
\newblock \emph{Applied ocean research}, 42:\penalty0 114--123, 2013.

\bibitem[Rezanejad et~al.(2013)Rezanejad, Bhattacharjee, and
  Soares]{rezanejad2013stepped}
K~Rezanejad, J~Bhattacharjee, and C~Guedes Soares.
\newblock Stepped sea bottom effects on the efficiency of nearshore oscillating
  water column device.
\newblock \emph{Ocean Engineering}, 70:\penalty0 25--38, 2013.

\bibitem[Rezanejad et~al.(2015)Rezanejad, Bhattacharjee, and
  Soares]{rezanejad2015analytical}
K~Rezanejad, J~Bhattacharjee, and C~Guedes Soares.
\newblock Analytical and numerical study of dual-chamber oscillating water
  columns on stepped bottom.
\newblock \emph{Renewable Energy}, 75:\penalty0 272--282, 2015.

\bibitem[Xu et~al.(2019)Xu, Wang, and Soares]{xu2019review}
Sheng Xu, Shan Wang, and C~Guedes Soares.
\newblock Review of mooring design for floating wave energy converters.
\newblock \emph{Renewable and Sustainable Energy Reviews}, 111:\penalty0
  595--621, 2019.

\bibitem[Khan and Behera(2020)]{khan2020analysis}
Mohamin Khan and Harekrushna Behera.
\newblock Analysis of wave action through multiple submerged porous structures.
\newblock \emph{Journal of Offshore Mechanics and Arctic Engineering},
  142\penalty0 (1), 2020.

\end{thebibliography}
%
%
%
%

\end{document}